%% file: final_paper.tex
\documentclass[preprint,12pt]{elsarticle}
\usepackage[utf8]{inputenc}
\usepackage[greek,english]{babel}
\usepackage{mathtools}
\usepackage{physics}
\usepackage{graphicx}
\newcommand{\en}{\selectlanguage{english}}
\usepackage{indentfirst}
\usepackage{tikz}
\usepackage{nicefrac}
\usepackage{wrapfig}
\usepackage{float}
\usepackage{tcolorbox}
\usepackage{afterpage}
\usepackage{amsmath}
\usepackage{amssymb} 
\usepackage[margin=1in]{geometry}
\usepackage{subcaption}
\usepackage{chngcntr}
\usepackage{empheq}
\usepackage{lineno}

\usepackage[obeyspaces]{url} %for paths
\usepackage{makecell}
\usepackage{booktabs}
\usepackage[unicode]{hyperref}
\hypersetup{colorlinks=true,linkcolor=black, urlcolor=black}
\usepackage{booktabs}
\usepackage{makecell}
\usepackage{float}
\usepackage[labelfont=bf]{caption}
\usepackage{enumitem}
\usepackage{tabularx}
\usepackage{booktabs}
\usepackage{epigraph}
\usepackage{csquotes}

\usepackage[numbers]{natbib}

\usepackage{orcidlink}

\usepackage[symbol]{footmisc}

\begin{document}
\begin{frontmatter}

\title{Search for an all-sky and a Galactic Ridge diffuse neutrino emission with the first 2 years of KM3NeT/ARCA data\\ (The KM3NeT Collaboration) }
% ----- Start automatically generated KM3NeT info
% ----- Start author list

\cortext[cor]{corresponding author}
\author[b,a]{O.~Adriani\,\orcidlink{0000-0002-3592-0654}}
\author[c,bf]{A.~Albert}
\author[d]{A.\,R.~Alhebsi\,\orcidlink{0009-0002-7320-7638}}
\author[d]{S.~Alshalloudi\,\orcidlink{0009-0000-6757-7224}}
\author[e]{F.~Ameli}
\author[f]{F.~Andersen}
\author[g]{M.~Andre}
\author[h]{L.~Aphecetche\,\orcidlink{0000-0001-7662-3878}}
\author[i]{M. Ardid\,\orcidlink{0000-0002-3199-594X}}
\author[i]{S. Ardid\,\orcidlink{0000-0003-4821-6655}}
% ...
% ----- End author list
% ----- Start address list
\address[a]{Universit{\`a} di Firenze, Dipartimento di Fisica e Astronomia, via Sansone 1, Sesto Fiorentino, 50019 Italy}
\address[b]{INFN, Sezione di Firenze, via Sansone 1, Sesto Fiorentino, 50019 Italy}
\address[c]{Universit{\'e}~de~Strasbourg,~CNRS,~IPHC~UMR~7178,~F-67000~Strasbourg,~France}
\address[d]{Khalifa University of Science and Technology, Department of Physics, PO Box 127788, Abu Dhabi,   United Arab Emirates}
\address[e]{INFN, Sezione di Roma, Piazzale Aldo Moro, 2 - c/o Dipartimento di Fisica, Edificio, G.Marconi, Roma, 00185 Italy}
\address[f]{IFIC - Instituto de F{\'\i}sica Corpuscular (CSIC - Universitat de Val{\`e}ncia), c/Catedr{\'a}tico Jos{\'e} Beltr{\'a}n, 2, 46980 Paterna, Valencia, Spain}
\address[g]{Universitat Polit{\`e}cnica de Catalunya, Laboratori d'Aplicacions Bioac{\'u}stiques, Centre Tecnol{\`o}gic de Vilanova i la Geltr{\'u}, Avda. Rambla Exposici{\'o}, s/n, Vilanova i la Geltr{\'u}, 08800 Spain}
\address[h]{Subatech, IMT Atlantique, IN2P3-CNRS, Nantes Universit{\'e}, 4 rue Alfred Kastler - La Chantrerie, Nantes, BP 20722 44307 France}
\address[i]{Universitat Polit{\`e}cnica de Val{\`e}ncia, Instituto de Investigaci{\'o}n para la Gesti{\'o}n Integrada de las Zonas Costeras, C/ Paranimf, 1, Gandia, 46730 Spain}
% ...
\address[bf]{Universit{\'e} de Haute Alsace, rue des Fr{\`e}res Lumi{\`e}re, 68093 Mulhouse Cedex, France}
% ...

%%%%%%%%%%%%%%%%%%%%%%%%%%%%%%%%%%%%%%%%%%%%%%%%%%%%%%%%%%%%%%%%%%%%%%%%%%%%%%%%%%%%%%%%%%%
%\linenumbers
\en
%%%%%%%%%%%%%%%%%%%%%%%%%%%%%%%%%%%%%%%%%%%%%%%%%%%%%%%%%%%%%%%%%%%%%%%%%%%%%%%%%%%%%%%%%%%
\newpage
\renewcommand{\abstractname}{{Abstract}}
\begin{abstract}
We report on the search performed for a diffuse astrophysical neutrino flux from the full sky and from  a specific region of the Galactic plane, the Galactic Ridge. The study uses the dataset collected with the first KM3NeT/ARCA configurations with 6, 8, 19, and 21 active detection units, corresponding to a total livetime of 640 days.
For the all-sky analysis, the fitted single-flavour astrophysical neutrino flux parameters, under the single power-law assumption, are $\phi_0^{1f} = 3.0^{+2.1}_{-2.0} \times 10^{-18}$ GeV$^{-1}$ cm$^{-2}$ s$^{-1}$ sr$^{-1}$ with spectral index $\gamma = 3.00^{+0.30}_{-0.35}$ at 68\% credible level. The Galactic Ridge fit does not yield constraints for the flux parameters with the current sensitivity.
For both analyses upper limits are derived and compared with state-of-the-art measurements.
While the analysis of these datasets has not yielded statistically significant results, the developed methods provide a solid basis for future measurements with the KM3NeT detector.
%While the analysis of these datasets has not produced statistically significant results, the developed methods establish a foundation  and demonstrate clear potential for guiding future measurements with the KM3NeT detector.
\end{abstract}
\end{frontmatter}

%%%%%%%%%%%%%%%%%%%%%%%%%%%%%%%%%%%%%%%%%%%%%%%%%%%%%%%%%%%%%%%%%%%%%%%%%%%%%%%%%%%%%%%%%%%

\newpage
\tableofcontents

%%%%%%%%%%%%%%%%%%%%%%%%%%%%%%%%%%%%%%%%%%%%%%%%%%%%%%%%%%%%%%%%%%%%%%%%%%%%%%%%%%%%%%%%%%%
\newpage
\section{Introduction}\label{sec:Introduction}

High-energy neutrinos produced in astrophysical environments propagate over cosmological distances while interacting weakly, and experiencing redshift energy losses and flavour oscillations. As electrically neutral particles, they are unaffected by magnetic fields, preserving directional information about the sources that generate them. In the standard scenario, cosmic-ray (CR) interactions with matter or radiation fields generate pions whose decay produces both neutrinos\footnote{Unless otherwise specified, the term neutrinos is used to denote both neutrinos and antineutrinos throughout this work.} and $\gamma$-rays. Therefore, neutrino flux predictions can be refined using cosmic-ray properties and $\gamma$-ray measurements, conducted by both space-based telescopes and ground-based detectors \cite{Halzen}.

One of the primary objectives of multi-messenger astronomy is to uncover the particle acceleration mechanisms operating in distant astrophysical sources. The measurement and characterisation of the cosmic neutrino flux offers valuable insights into these processes \cite{Ackermann}.

Astrophysical neutrinos can be detected by large-scale neutrino telescopes through the Cherenkov radiation induced by secondary charged particles produced in neutrino interactions within or near the detector. These instruments are optimised for TeV$-$PeV energies, where atmospheric backgrounds are strongly suppressed \cite{Spurio}.

%-------------------------------------------
\subsection{Diffuse all-sky emission}
A high-energy diffuse flux of cosmic neutrinos should arise from the combined contributions of unresolved individual neutrino sources and from hadronic interactions during  cosmic-ray propagation.

The energy spectrum of astrophysical neutrinos is commonly modelled by a power law with single-flavour normalisation $\phi_0$ and spectral index $\gamma$ :

\begin{equation}\label{eq:diffuse flux standard equation}
\Phi_{\nu + \bar{\nu}}(E) = \frac{dN}{dE} = \phi_0 \cdot 10^{-18} \left(\frac{E}{100\, \mathrm{TeV}}\right)^{-\gamma} \quad
\left[\mathrm{GeV}^{-1} \mathrm{cm}^{-2} \mathrm{s}^{-1} \mathrm{sr}^{-1}\right].
\end{equation}
%where \(\phi_0\) and \(\gamma\) denote the single-flavour normalisation and spectral index, respectively.

The first compelling evidence for a diffuse flux of astrophysical neutrinos was reported in 2013 by the IceCube Collaboration~\cite{IceCube2013}. This High-Energy Starting Events (HESE) analysis, based on neutrino interactions contained within the detector volume, demonstrated sensitivity to all flavours and arrival directions while efficiently suppressing atmospheric backgrounds. Subsequent updates with larger exposure and improved systematic treatments strengthened the rejection of the background-only hypothesis and improved the flux characterisation~\cite{IceCube2020hese}.
Independent IceCube analyses based on different datasets, including a variety of event topologies and selections~\cite{IceCube2021_muontracks, IceCube_cascade, Diffuse_IceCube2024} have consistently confirmed the existence of a diffuse astrophysical component over an energy range spanning from a few TeV to several PeV.
The Baikal-GVD Collaboration has also recently reported a significant observation of the diffuse flux~\cite{Baikal_GVDdiffuse}, while the ANTARES Collaboration has provided constraints on the properties of the cosmic neutrino spectrum~\cite{ANTARES_diffuse_2024}.

Despite the isotropic single power-law assumption, discrepancies between different measurements suggest possible deviations from a single-component scenario~\cite{PalladinoVissaniSpurio}. This has motivated interpretations involving multiple components, such as, for example, a Galactic contribution superimposed on an extragalactic flux. The recent observation of a diffuse neutrino emission from the Galactic plane~\cite{IC_GP} provides qualitative support for this hypothesis, with preliminary estimates placing the Galactic contribution at the level of $\sim$ 6\% $-$ 13\% at 30 TeV. However, its precise magnitude remains uncertain due to differences in the assumed energy ranges and spectral shapes used in template-based analyses. A robust assessment of the prompt atmospheric neutrino component is also essential to fully clarify the nature of the observed diffuse flux~\cite{Prompt_importance}.

Located in the Northern Hemisphere, the KM3NeT detectors can observe the centre of the Galaxy with sub-degree angular precision, enabling detailed studies and precise measurements of neutrino emissions and of the mechanisms responsible for Galactic particle acceleration.
%-------------------------------------------
\subsection{Galactic Ridge emission}

The Galactic plane is the most prominent source in the sky across all electromagnetic wavelengths, and is expected to contribute to the diffuse astrophysical neutrino flux. 
In recent years, several theoretical models have been developed to constrain the flux of neutrinos originating from the Galactic plane~\cite{cragamma,CRINGE,Evoli}.  Specifically, the predicted neutrino flux is comparable in magnitude to the diffuse $\gamma$-ray flux observed by Fermi-LAT~\cite{FermiLAT_2012}. 

In the inner region of the Galactic plane, known as the Galactic Ridge (defined by Galactic longitude $l$ and Galactic latitude $b$, with $|l|$ $<$ 30$^{\circ}$ and $|b|$ $<$ 2$^{\circ}$), the diffuse $\gamma$-ray measurements suggest a harder CR spectral slope than that measured locally at Earth~\cite{FermiLAT_2012,TibetASgamma,LHAASO}. This region encompasses the Galactic bar and the innermost parts of the spiral arms, which are responsible for the highest star formation rates in our Galaxy.

In recent years, several experiments have reported results on the search for a neutrino flux from the Galactic plane. The ANTARES Collaboration found an excess of events originating from the Galactic Ridge, incompatible with the background-only hypothesis at the 96$\%$ confidence level~\cite{ANTARES_gp}.
The IceCube Collaboration reported the first observation of high-energy neutrino emission from the Galactic plane at a significance level of 4.5$\sigma$~\cite{IC_GP}. The detected signal is consistent with neutrino production via CR interactions with gas in the interstellar medium, corroborating the Milky Way as a source of high-energy neutrinos and providing constraints on Galactic CR acceleration and transport models.

\subsection{Objectives} 
The primary objective of the all-sky and Galactic Ridge analyses presented in this paper is to assess a potential excess of high-energy events independently in the two analyses, and determine the best-fit values for the signal flux, and the relative contribution of the Galactic neutrino emission to the overall cosmic diffuse flux. 
%The cosmic signal consists of neutrinos detectable via both charged-current and neutral-current interactions.

The main difference between the methodologies in the two analyses lies in the background determination. In the all-sky diffuse flux analysis, Monte Carlo simulations are used to estimate the expected background rate after the final selection. In contrast, the Galactic Ridge analysis exploits the KM3NeT telescope’s location and the smaller solid angle subtended by the Galactic Ridge, directly determining the background from data by considering events from sky regions with equal exposure where no signal is expected.\\

The structure of this paper is as follows. The detector and the neutrino data samples used in the analysis, including details of the event selection procedure, are described in Sections~\ref{sec:detector} and~\ref{sec:dataset}. The adopted statistical framework and the evaluation of systematic uncertainties are outlined in Section~\ref{sec:systematics}. The results for the all-sky diffuse flux and for the Galactic Ridge are presented in Sections~\ref{sec:all_sky} and~\ref{sec:GR}, respectively.
Conclusions and perspectives are provided in Section~\ref{sec:conclusions}.

\section{The KM3NeT detector}\label{sec:detector}

The KM3NeT (Cubic Kilometre Neutrino Telescope) Collaboration is constructing a network of second-generation neutrino telescopes on the seabed of the Mediterranean Sea, a region offering ideal conditions for operating Cherenkov detectors at abyssal depths~\cite{LetterIntent}. KM3NeT  profits from the experience gained from ANTARES~\cite{Antares_legacy}, which operated successfully for over 15 years before being switched off in February 2022.

The KM3NeT/ARCA (Astroparticle Research with Cosmics in the Abyss) detector is located approximately 100 km off the coast of Sicily at a depth of 3500 metres and is optimised for the study and identification of astrophysical neutrino fluxes, targeting the TeV$-$PeV energy range~\cite{Thijs}. Once fully constructed, it will instrument over one cubic kilometre of seawater.
The KM3NeT/ORCA (Oscillation Research with Cosmics in the Abyss) detector is being constructed off the coast of Toulon, France, at a depth of 2500 metres, close to the ANTARES deployment site, and aims at resolving the neutrino mass ordering by detecting atmospheric neutrinos in the few GeV$-$TeV range~\cite{Orca6}. Its final configuration will encompass around 7 megatons of seawater.

Both KM3NeT/ORCA and KM3NeT/ARCA are based on the same construction technology, namely a three-dimensional array of optical sensors, and rely on the same detection principle.
Each Detection Unit (DU) is a vertical string-like structure anchored to the seabed comprising 18 Digital Optical Modules (DOMs)~\cite{DOM}. A DOM constitutes the fundamental detection element and houses 31 3-inch photomultiplier tubes (PMTs). The primary distinction between the two configurations lies in the vertical and horizontal spacing of DOMs and DUs to match their targeted energy ranges. In KM3NeT/ARCA (ORCA), DOMs are separated by approximately 36 m (9 m) along a DU, with an average inter-DU spacing of about 90 m (20 m).
The analyses presented in this article focus on the KM3NeT/ARCA detector, while it was taking data with 6 to 21 DUs, see Table~\ref{tab:ARCA periods}. 

\subsection{Event reconstruction} 

The KM3NeT neutrino telescope employs multiple reconstruction algorithms to infer the properties of neutrino interactions, such as the arrival direction of the parent neutrino and deposited energy. These algorithms are optimised for two distinct event topologies observed in the detector, namely \textit{track-like} and \textit{shower-like} signatures.

The event topology is primarily determined by the interaction channel and neutrino flavour. Charged-current $\nu_\mu$ interactions produce long muon tracks, whose direction can be reconstructed with sub-degree angular resolution. In contrast, charged-current interactions of $\nu_e$ and $\nu_\tau$, as well as neutral-current interactions of all flavours, predominantly yield shower-like topologies. When fully contained within the instrumented volume, these events allow for a more accurate estimation of the deposited energy, albeit with a comparatively reduced angular resolution~\cite{KM3NeT_reco}.

The analyses presented in this article are focused on the selection of track-like neutrino events. In the following paragraph the track reconstruction algorithm is described~\cite{Thijs,Nature}.

\subsubsection{Track reconstruction}\label{sec:track_reco}
The KM3NeT neutrino telescope employs multi-stage track reconstruction algorithms to determine the direction and energy of ultra-relativistic muons produced in charged-current $\nu_\mu$ interactions, exploiting the Cherenkov light emitted as the muons traverse the detector. Reconstruction begins with a prefit stage, in which the muon is approximated as a straight line and expected arrival times of photons on the PMTs are computed using a simplified model that neglects light scattering and dispersion. 
Causally connected hit\footnote{A hit is defined as the integrated electric charge of the PMT signal over a certain threshold, the crossing time and the duration of the signal (Time over Threshold).} clusters are selected to suppress optical background and scattered photons, and the procedure is repeated over a grid of directions with 1$^{\circ}$ spacing. In KM3NeT/ARCA, the 50 best prefit solutions are retained as seeds for the subsequent fit. These seeds initialise a maximum-likelihood muon trajectory fit that incorporates light propagation effects in seawater, such as scattering and absorption. The likelihood is constructed from the time-residual probability density function of the first detected hit on each PMT \cite{mdj}, thereby ensuring robustness against afterpulses and modelling uncertainties. The fit iteratively refines the track parameters (two directional angles, time, and spatial coordinates) using only hits within a cylindrical volume aligned with the prefit direction. The muon energy is estimated at the final reconstruction stage from the energy loss per unit track length, based on the detected hits by the PMTs, the reconstructed track length within the instrumented volume, and the individual PMT detection efficiencies. The reconstruction algorithm returns quality and performance parameters, including the track likelihood and an estimate of the arrival direction angular uncertainty ($\beta_0$), which are subsequently used for event selection.

%%%%%%%%%%%%%%%%%%%%%%%%%%%%%%%%%%%%%%%%%%%%%%%%%%%%%%%%%%%%%%%%%%%%%%%%%%%%%%%%%%%%%%%%%%%
\section{Dataset and event selection}\label{sec:dataset}

The modular architecture of the KM3NeT detector enables data acquisition even during construction, with only partial configurations in place. Construction proceeds in phases, with a variable number of DUs deployed in each sea campaign. Data-taking periods are defined by the number of DUs active at a given time, with stable data collection starting once the newly deployed DUs have been commissioned.

The DUs are subject to movement due to sea currents. A dedicated calibration system, consisting of an acoustic positioning system combined with gyroscopes and additional sensors embedded in each DOM, ensures sub-degree angular resolution by tracking the position of every active element over time with an accuracy of better than 20~cm~\cite{KM3NeT_calibration}.

The data analysed in this work correspond to detector configurations with different numbers of DUs, denoted by the number following the acronym ARCA, as reported in Table~\ref{tab:ARCA periods} together with the corresponding livetime.

\begin{table}[htbp]
\centering
\caption{\footnotesize KM3NeT/ARCA configurations and analysed livetime.}
\begin{tabularx}{\textwidth}{p{3cm} >{\centering\arraybackslash}p{8cm} >{\centering\arraybackslash}p{3cm}}
\toprule
\toprule
\multicolumn{1}{l}{\textbf{Name}} & \multicolumn{1}{c}{\textbf{Period}} & \multicolumn{1}{c}{\textbf{Livetime (days)}} \\
\midrule

ARCA6 & May 2021 --- September 2021 & 92.1 \\
ARCA8 & September 2021 --- June 2022 & 212.3 \\
ARCA19 & July 2022 --- September 2022 & 48.4 \\
ARCA21 & September 2022 --- September 2023 & 287.4 \\

\bottomrule
\bottomrule
\end{tabularx}
\label{tab:ARCA periods}
\end{table}

The recorded data stream is dominated by background events arising from atmospheric muons and atmospheric neutrinos reaching the detector.
The analyses described here require a high-purity neutrino selection, which is achieved through a series of cuts on variables resulting from the track reconstruction algorithm. To further improve the purity of the sample, machine learning techniques are applied, specifically a Boosted Decision Tree (BDT) classifier, see Section~\ref{sec:Boosted Decision Tree}.

Monte Carlo simulations are used to model the detector response, validate the analysis methodology, and fine-tune background rejection strategies.
Atmospheric muons are simulated using the \texttt{MUPAGE} software~\cite{mupage}. Neutrino interactions are simulated with the \texttt{gSeaGen} code~\cite{MC_km3_1,MC_km3_2}. 

The atmospheric neutrino flux comprises two components: the \emph{conventional} flux, produced by the decay of light, long-lived mesons, dominating up to energies of 10$-$100~TeV is described by the Honda et al.\ model~\cite{Honda_2006}; the \emph{prompt} flux, originating from the decay of heavier, charmed short-lived hadrons and described according to the Enberg et al. model~\cite{Enberg}. These fluxes are further corrected to shape the cosmic-ray knee using the H3a composition model~\cite{Gaisser_2013}.

A blind analysis strategy is employed to mitigate selection bias in the optimisation procedure. All event selection criteria, including the BDT training and optimisation as well as subsequent cut tuning, are derived exclusively from Monte Carlo simulations. A 10\% of the recorded data sample is used as a control sample. The full dataset is used after unblinding for the final analysis. %is \textit{unblinded} and the predefined selection is applied without further modification.

\subsection{Preliminary event selection}
\label{sec:Preselection}

\begin{figure}[!ht]
    \centering
    \begin{minipage}[c]{0.49\textwidth}
    \includegraphics[width=1.\textwidth]{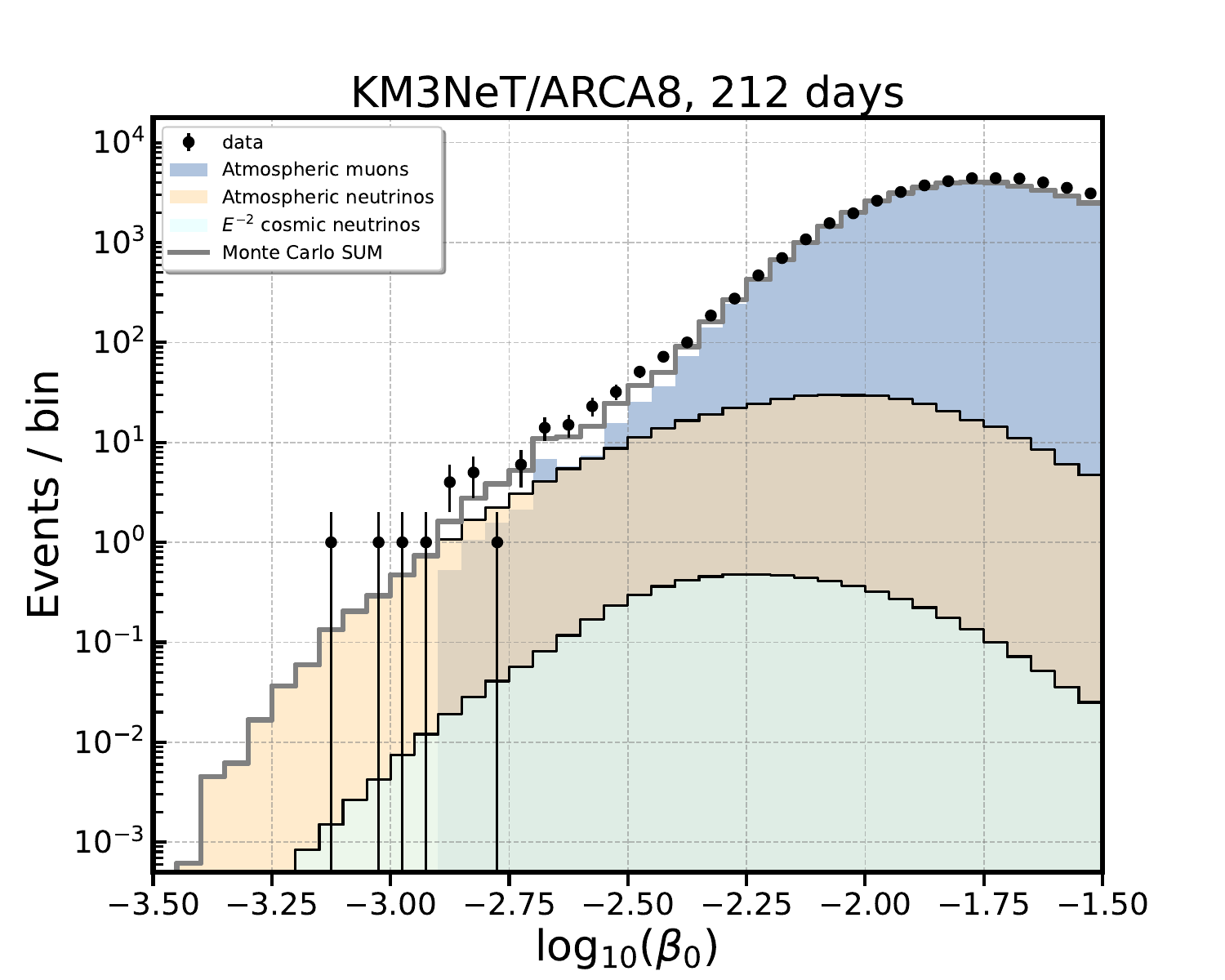}
    \end{minipage}
    \begin{minipage}[c]{0.49\textwidth}
    \includegraphics[width=1.\textwidth]{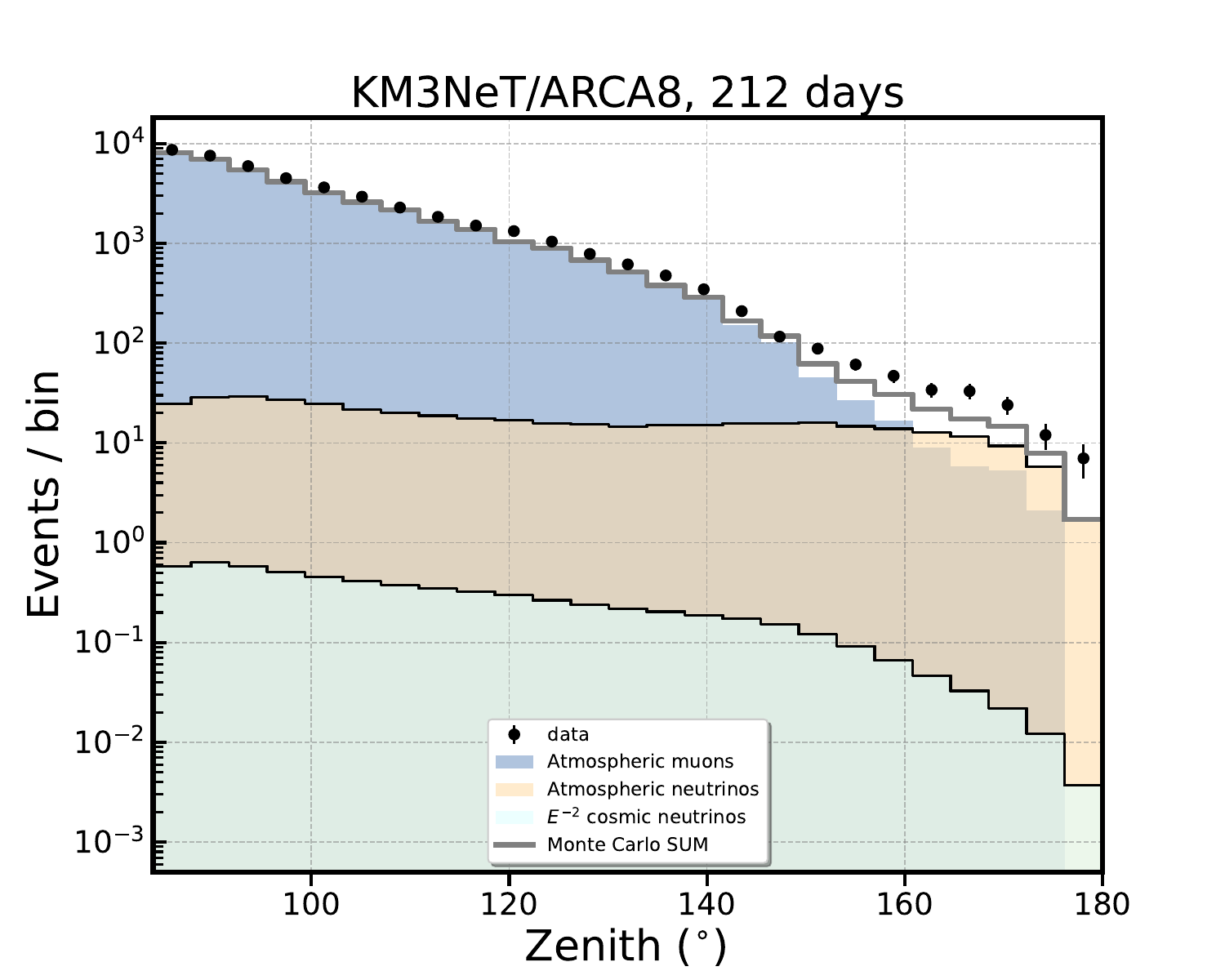}
    \end{minipage}
    \caption{\footnotesize Data and Monte Carlo comparison for the angular error estimate ($\beta_0$) (left) and the reconstructed zenith angle (right) after preliminary event selection for the ARCA8 configuration. Statistical uncertainties are shown for data.}
    \label{fig:ARCA8_beta0_zenith}
    \end{figure}

\begin{figure}[!ht]
    \centering
    \begin{minipage}[c]{0.49\textwidth}
    \includegraphics[width=1.\textwidth]{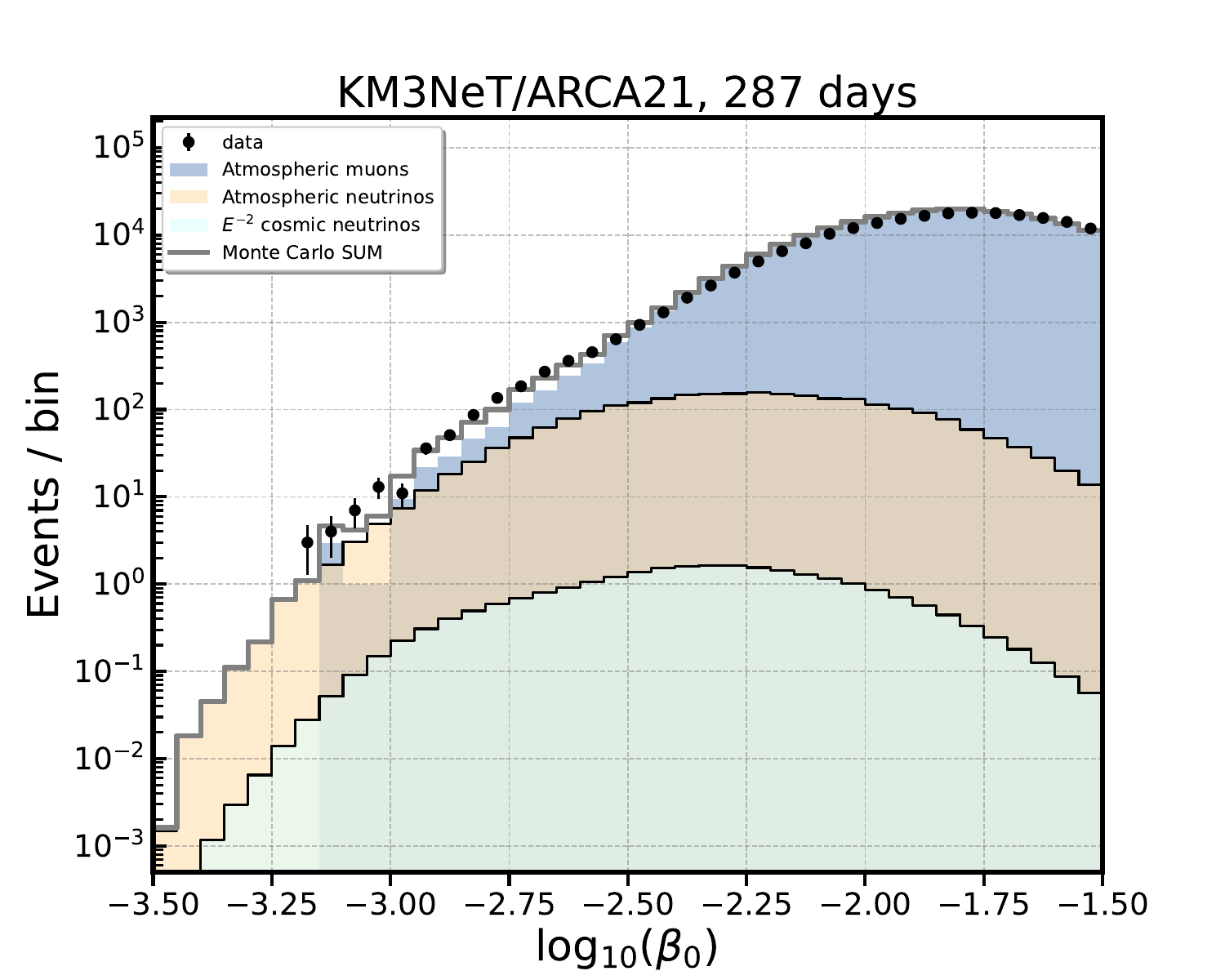}
    \end{minipage}
    \begin{minipage}[c]{0.49\textwidth}
    \includegraphics[width=1.\textwidth]{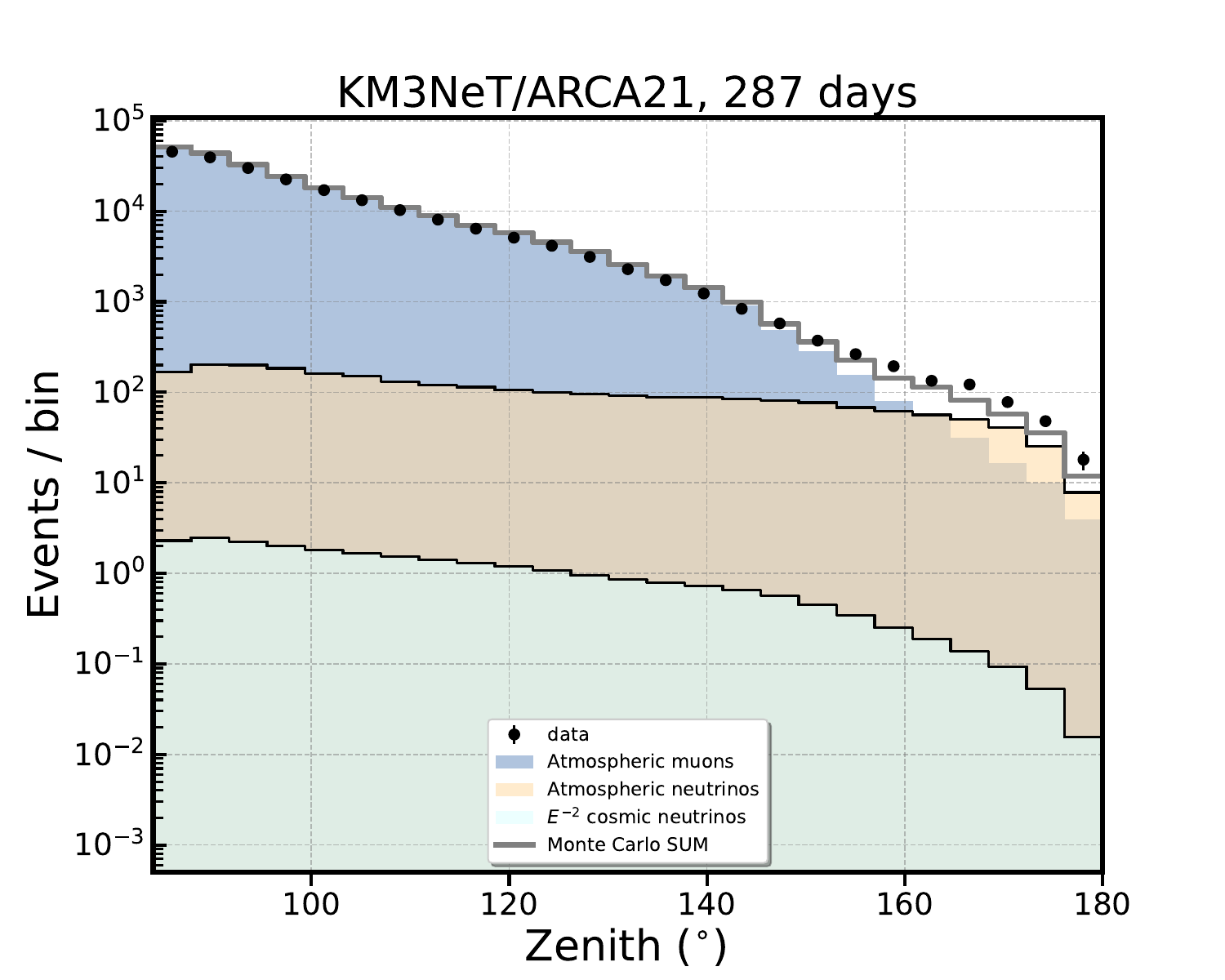}
    \end{minipage}
    \caption{\footnotesize Data and Monte Carlo comparison for the angular error estimate ($\beta_0$) (left) and the reconstructed zenith angle (right) after preliminary event selection for the ARCA21 configuration. Statistical uncertainties are shown for data.}
    \label{fig:ARCA21_beta0_zenith}
    \end{figure}

The preliminary event selection begins with the suppression of events arising from random coincidences of noise hits on the optical modules,  due to the radioactive decays in seawater and bioluminescent activity in the underwater environment.
To effectively mitigate this background component, trigger-level information is combined with track-reconstruction quality variables, leaving atmospheric muons as the dominant residual background. The atmospheric muon event rate, which exceeds the one from neutrino interactions by several orders of magnitude, can be significantly reduced by selecting tracks with a reconstructed zenith angle below the horizon, as atmospheric muons cannot traverse the Earth. At this stage, a requirement is imposed by selecting only events whose reconstructed zenith angle is greater than 84$^\circ$. Additional constraints are applied to the maximised log likelihood (see Section~\ref{sec:track_reco}), length of the reconstructed track, as well as to the angular uncertainty in the track reconstructed direction ($\log_{10}(\beta_0$)). This selection serves as a unified starting point for both analyses. For further details see Sections~\ref{sec:all_sky} and~\ref{sec:gr_optimisation}. Figures~\ref{fig:ARCA8_beta0_zenith} and~\ref{fig:ARCA21_beta0_zenith} show the comparison of representative variables for data and Monte Carlo.

\newpage
\subsection{Boosted Decision Tree}
\label{sec:Boosted Decision Tree}

Although the application of the preliminary cuts results in a significant reduction of the atmospheric muon background, the residual rate still exceeds the expected cosmic neutrino signal by several orders of magnitude. Since atmospheric muons cannot traverse the Earth, the remaining events are mostly down-going particles misreconstructed as up-going. These events are rejected through a Boosted Decision Tree classifier (more details in~\cite{Vasilis_thesis}) trained with the ROOT TMVA package~\cite{tmva}.

Two separate BDT models, one trained for the ARCA6/ARCA8 and one for the ARCA19/ ARCA21 detector configurations, were optimised to maximise the rejection of atmospheric muon background while preserving signal efficiency.
In order to select a subset of variables from the full set of available inputs, their discriminating power as a BDT feature has been evaluated with the {\em SelectKBest} tool from the \texttt{Scikit-learn} library \cite{scikit-learn}, with the ANalysis Of VAriance (ANOVA) F-test as scoring function. Features with higher F-values provide better class separation and were selected as the most discriminative inputs for the analysis. The architecture of the BDT model is reported in~\ref{appendix:bdt}.

\begin{figure}[!ht]
\centering
  \includegraphics[width=0.7\textwidth]{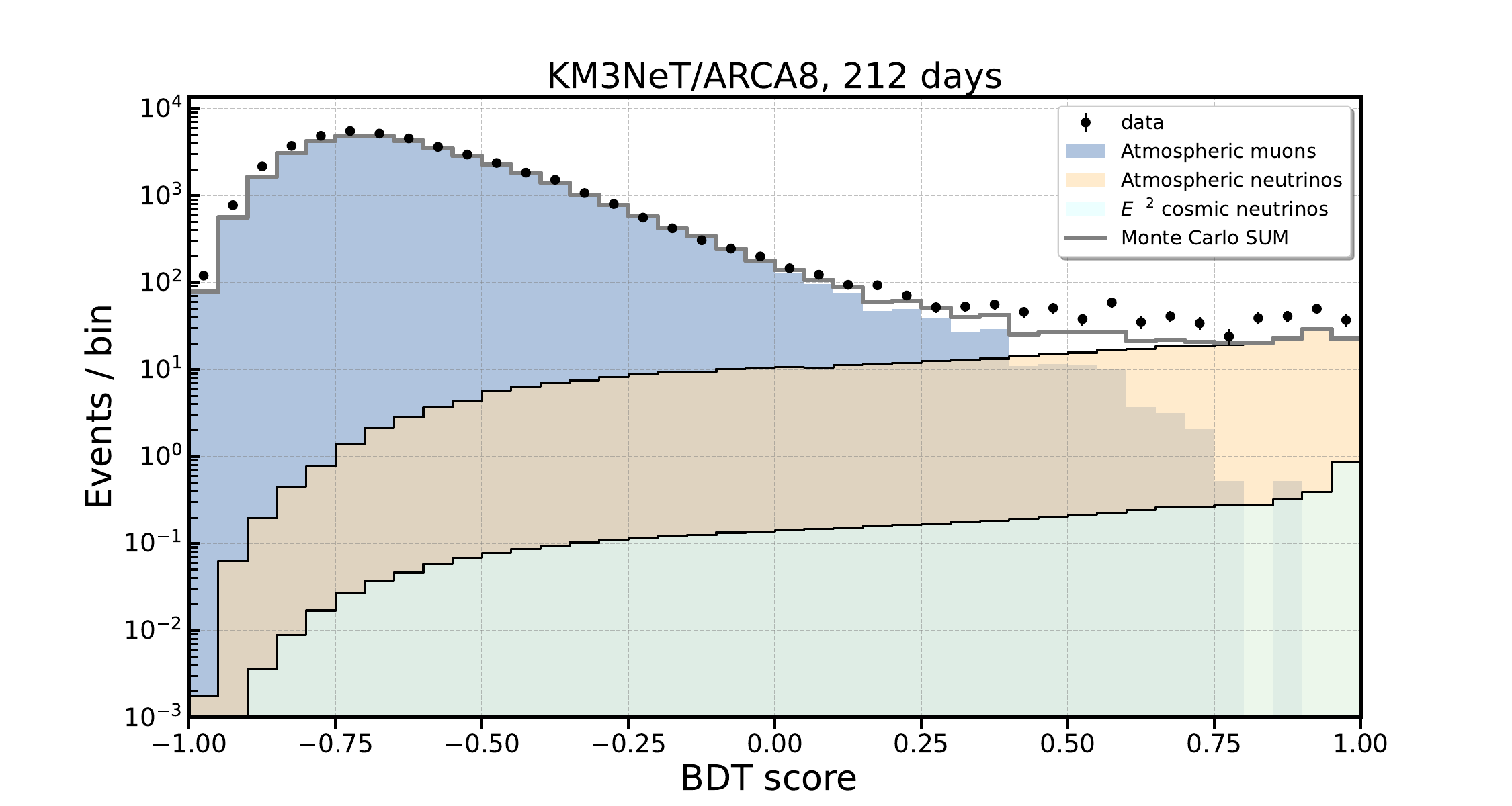}
  \includegraphics[width=0.7\textwidth]{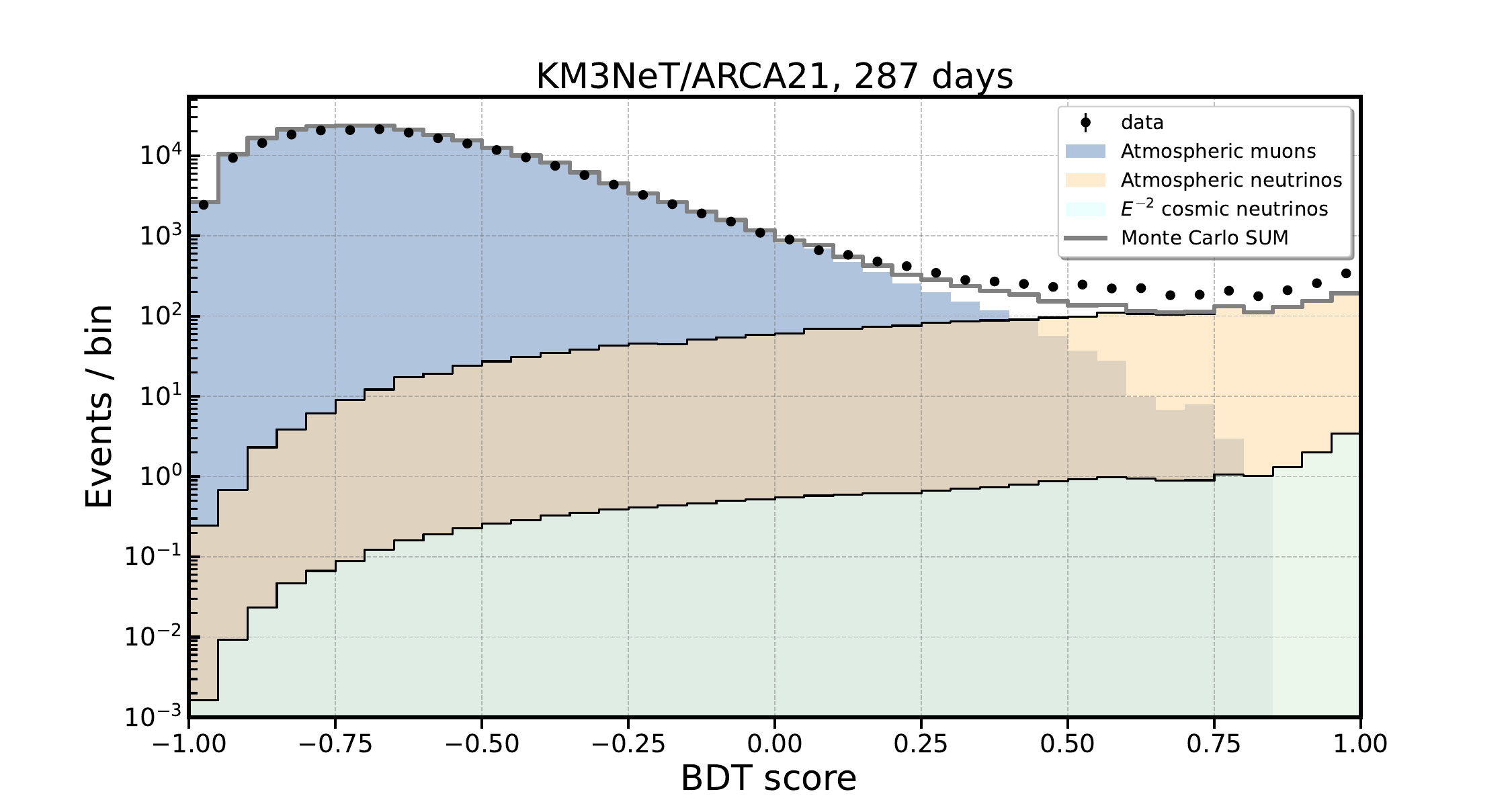}
  \caption{\footnotesize BDT score distributions for data (black dots with statistical errors) and Monte Carlo simulated events for ARCA8 (top) and ARCA21 (bottom): atmospheric muons (blue), atmospheric neutrinos (orange) and cosmic neutrinos (green).}
  \label{fig:BDT_scores}
\end{figure}

The optimal set of hyperparameters for the BDT is determined using a Grid Search technique based on a custom framework~\cite{Stavropoulos2022}. In this approach, $\sim$ 2500 models were trained with different hyperparameter combinations, each evaluated at specific BDT score thresholds and ranked according to several performance metrics, such as signal efficiency, F1-score \cite{f1_score}, and data/MC agreement.
Following the blinding policy already described, 10\% of the data for each configuration was used for the evaluation of the models. The BDT score distributions for MC simulated events are shown in Figure~\ref{fig:BDT_scores}, overlaid with data after {\em unblinding}.
The optimisation of the final event selection is carried out separately for the two analyses, each following its own background estimation approach and methodological framework. In both cases, the Model Rejection Factor (MRF) technique~\cite{MRF} is used to define the optimal set of cuts that maximise the sensitivity to a given signal flux.

%%%%%%%%%%%%%%%%%%%%%%%%%%%%%%%%%%%%%%%%%%%%%%%%%%%%%%%%%%%%%%%%

\section{Statistical analysis and systematic uncertainties}\label{sec:systematics}
A Bayesian approach is used to analyse data and to fit the parameters of interest for both analyses, following the method developed in \cite{ANTARES_gp}, while a frequentist analysis is also presented in Section~\ref{sec:frq_results}.
The likelihood function is expressed as the product of the Poisson probabilities evaluated for each bin $i$ to observe $N_i$ events, given an expected background $B_i$ and a signal contribution $\phi_0 S_i(\gamma)$ computed for spectral index $\gamma$ and normalisation $\phi_0$:

\begin{equation}\label{eq:likelihood}
L(\{N_i\}; \{S_i(\gamma)\}, \{B_i\}, \phi_0) = \prod_i \, \text{Poisson}(N_i, \, B_i + \phi_0 S_i(\gamma)).
\end{equation}

The posterior probability function is then obtained incorporating Gaussian priors to account for background statistical and systematic uncertainties, $\pi(\{B_i\})$, and for signal acceptance, $\pi(\{S_i(\gamma)\})$. A flat prior on the signal spectral shape is considered $\pi(\phi_0, \gamma)$, since no previous knowledge of $\phi_0$ and $\gamma$ is assumed.
The posterior distribution is then marginalised over the nuisance parameters:

\begin{align}\label{eq:marginalised posterior}
p(\phi_0, \gamma; \{N_i\}) = \int &L(\{N_i\}; \{S_i(\gamma)\}, \{B_i\}, \phi_0) \times\\
&\times \pi(\{B_i\}) \times \pi(\{S_i(\gamma)\}) \times \pi(\phi_0, \gamma) \times  \prod_i \times dB_i \, dS_i(\gamma).\notag
\end{align}

Different detector configurations can be combined irrespective of the binning scheme or energy range by multiplying the likelihoods corresponding to each configuration and then computing the marginalised posterior.

\begin{figure}[H]
	\centering	
	\includegraphics[width=0.75\textwidth]{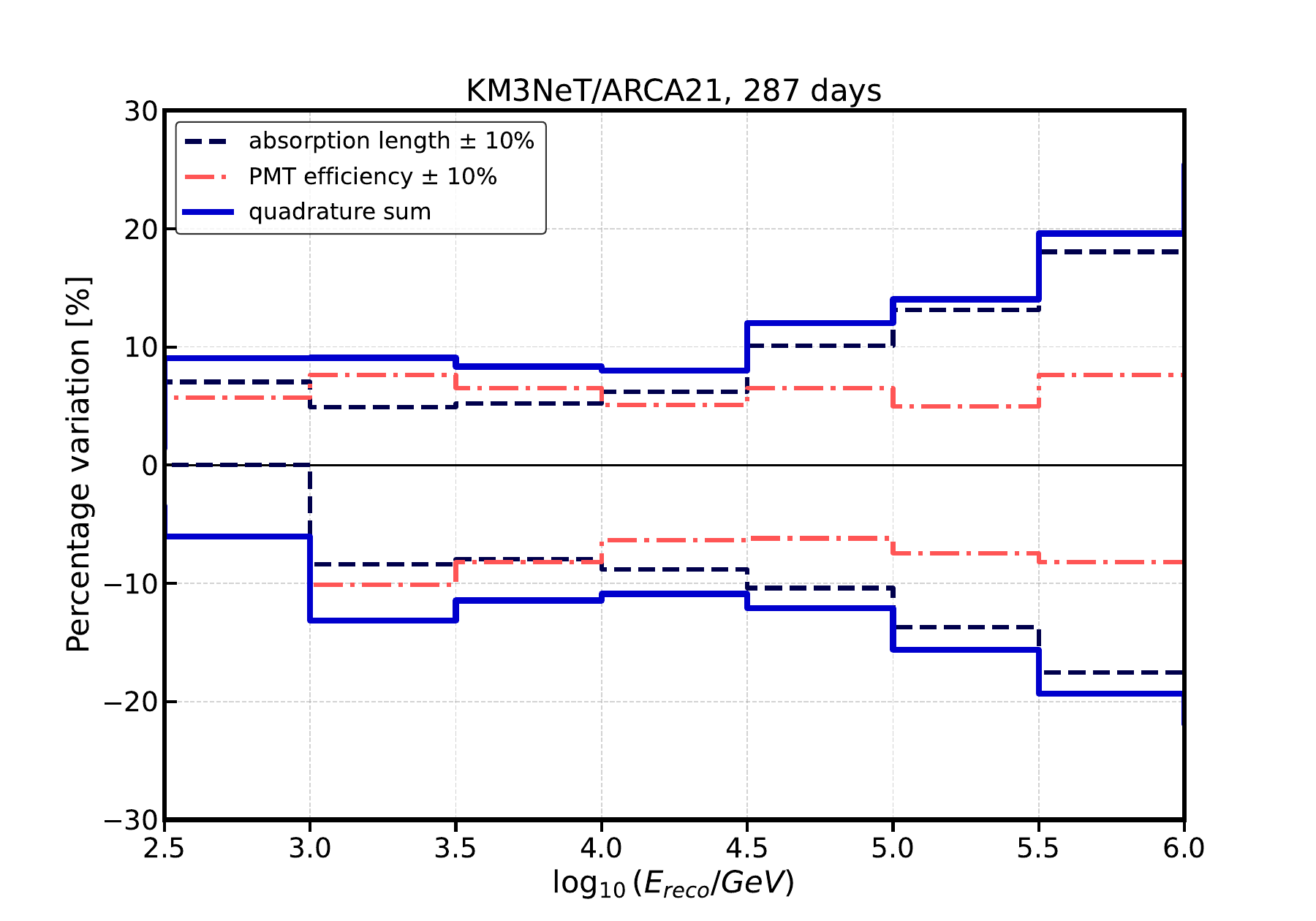}
	\caption{\footnotesize Percentage variation relative to the nominal simulation as a function of reconstructed neutrino energy for an $E^{-2}$ astrophysical neutrino flux. It was obtained varying PMT efficiency and light absorption length from modified Monte Carlo simulations. The quadrature sum is taken as the uncertainty on the signal acceptance.}
	\label{Systematics}
\end{figure}

The marginalised posterior distribution in Eq.~\ref{eq:marginalised posterior} is used to estimate: \textit{i)} the {\em upper limit} (UL) on the flux normalisation $\phi_0^{UL}$ with respect to a specific spectral index profiling the posterior probability for a selected $\gamma$, \textit{ii)} the {\em sensitivity} to a flux computed as the median UL for a collection of background-only pseudo-experiments and, \textit{iii)} the {\em best-fit} values of $\phi_0$ and $\gamma$, corresponding to the maximum of the posterior distribution. The systematic errors included as priors in the analyses have been evaluated using dedicated Monte Carlo simulations, with a limited number of parameters identified as the main contributors. To differentiate and quantify their impact, specific simulations have been performed varying one
parameter at a time. For both analyses the following sources of systematic uncertainties were considered, assumed to be uncorrelated:
\begin{itemize}
    \item PMT efficiency. The efficiency is varied by a $\pm 10\%$.
    \item Light absorption length in water. An uncertainty of $\pm 10\%$ is considered based on existing measurements performed in the southern Mediterranean Sea~\cite{Riccobene, Nestor2, NEMO_optical}. This value is consistent within uncertainties with measurements performed at the ANTARES site~\cite{Aguilar2010}. 
\end{itemize}
The results, obtained from neutrino simulations weighted according to an $E^{-2}$ flux and expressed as the percentage variation in the number of events with respect to the nominal simulation, are shown in Figure~\ref{Systematics}. The resulting variation is taken as an estimate of the systematic uncertainty on the signal acceptance and is incorporated consistently in both analyses. Variations in the PMT efficiency and in the light absorption length lead to a consistent increase or decrease in the number of triggered events. These effects predominantly impact low-energy events, which are reconstructed across a broad region of the phase space. The overall variation is consistent with previous measurements performed by ANTARES~\cite{ANTARES_QE}.
In the all-sky diffuse flux analysis, where the background is obtained from Monte Carlo simulations, an additional systematic uncertainty of $\pm$40\% that takes into account the normalisation of the atmospheric muon and neutrino flux~\cite{Honda_2006} and detector acceptance~\cite{ANTARES_QE} is considered. All contributions are combined in quadrature resulting in a total background systematic uncertainty of 40\% and 45\% for ARCA6$-$8 and ARCA19$-$21, respectively.

%%%%%%%%%%%%%%%%%%%%%%%%%%%%%%%%%%%%%%%%%%%%%%%%%%%%%%%%%%%%%%%%

\section{All-sky diffuse neutrino flux}
\label{sec:all_sky}
For the search of a diffuse all-sky neutrino flux only up-going (zenith angle $>$ 90$^{\circ}$) reconstructed events are considered. Additional constraints on the maximised log likelihood and on the number of triggered DOMs are applied to reduce the contamination due to badly reconstructed atmospheric muons.
%and thus the recently observed event KM3-230213A~\cite{Nature} is not included in the sample.
For the subsequent optimisations, following Eq.~\ref{eq:diffuse flux standard equation}, the parameters measured by IceCube $\phi_{0} =$ 1.44 $\times$ 10$^{-18}$ GeV$^{-1}$~cm$^{-2}$~s$^{-1}$~sr$^{-1}$ and $\gamma = 2.37$~\cite{IceCube2021_muontracks} are adopted as a baseline.
The lowest MRF identifies both the optimal BDT and reconstructed energy cuts, as reported in Table~\ref{tab:bdt_score_cuts_allsky}.

\begin{table}[H]
\centering
\caption{\footnotesize Optimal set of requirements on BDT score and reconstructed energy, $\log_{10}$ ($E_{\text{reco}}$/GeV), found for each ARCA geometry through the MRF procedure.}
\begin{tabular}{lcc}
\toprule
\toprule
\textbf{ARCA configuration} & \textbf{BDT Score} & \textbf{log$_{10}$ (\textit{E}$_{\text{reco}}$/GeV)} \\
\hline
ARCA6  & $> 0.35$ & $> 4.00$ \\
ARCA8  & $> 0.27$ & $> 4.20$\\
ARCA19 & $> 0.45$ & $> 4.20$ \\
ARCA21 & $> 0.40$ & $> 4.36$ \\
\bottomrule
\bottomrule
\end{tabular}
\label{tab:bdt_score_cuts_allsky}
\end{table}

The poor statistics of high-energy atmospheric muons at high BDT scores is mitigated by applying a Gaussian extrapolation fit to the reconstructed muon energy distribution after imposing the BDT score cut.
For most of the selected data the purity of the neutrino sample of about 97\%  is estimated.
For the ARCA6 configuration, the data sample requires a tighter selection on the variable $\log_{10}(\beta_0$) following the BDT score cut to further reduce the atmospheric muon background.

\subsection{Detector performance}
To obtain a quantitative estimate of the selection efficiency, the number of expected events and the observed data for each period after the optimal BDT score cut, are summarised in Table~\ref{tab:Events before and after BDT}. 
\begin{table}[H]
\centering
\caption{\footnotesize Number of events for the different ARCA configurations after the optimal BDT score cut obtained for the all-sky diffuse flux sample, in the range $\log_{10} (E_{\rm reco}/\text{GeV}) \in [2,8]$. The expected number of cosmic neutrinos has been derived for the flux parameters $\phi_{0} =$ 1.44 $\times$ 10$^{-18}$ GeV$^{-1}$ cm$^{-2}$ s$^{-1}$ sr$^{-1}$ and $\gamma = 2.37$~\cite{IceCube2021_muontracks}.}
\setlength{\tabcolsep}{8pt} 
\renewcommand{\arraystretch}{1.4} % row spacing
\begin{tabular}{l|cccc}%{l|c|c|c|c}
\toprule
\toprule
 & \textbf{ARCA6} & \textbf{ARCA8} & \textbf{ARCA19} & \textbf{ARCA21} \\
\midrule
Atm.~$\nu$ (conv. + prompt) & 70.2 & 174.2 & 151.1 & 958.8 \\
Cosmic~$\nu$ & 1.4 & 4.0 & 2.5 & 16.3 \\
Atm. $\mu$ & 32.7  & 31.7 &  6.3 & 29.9 \\
Data & 111 & 340 & 232 & 1791 \\
\bottomrule
\bottomrule
\end{tabular}
\label{tab:Events before and after BDT}
\end{table}

The discrepancy between the data and Monte Carlo predictions is addressed by introducing a free parameter in the fit, allowing for the background normalisation to vary.
This approach accounts for differences between simulation and data, thereby reducing potential biases in the final result.

\begin{figure}[H]
  \includegraphics[width=0.5\textwidth]{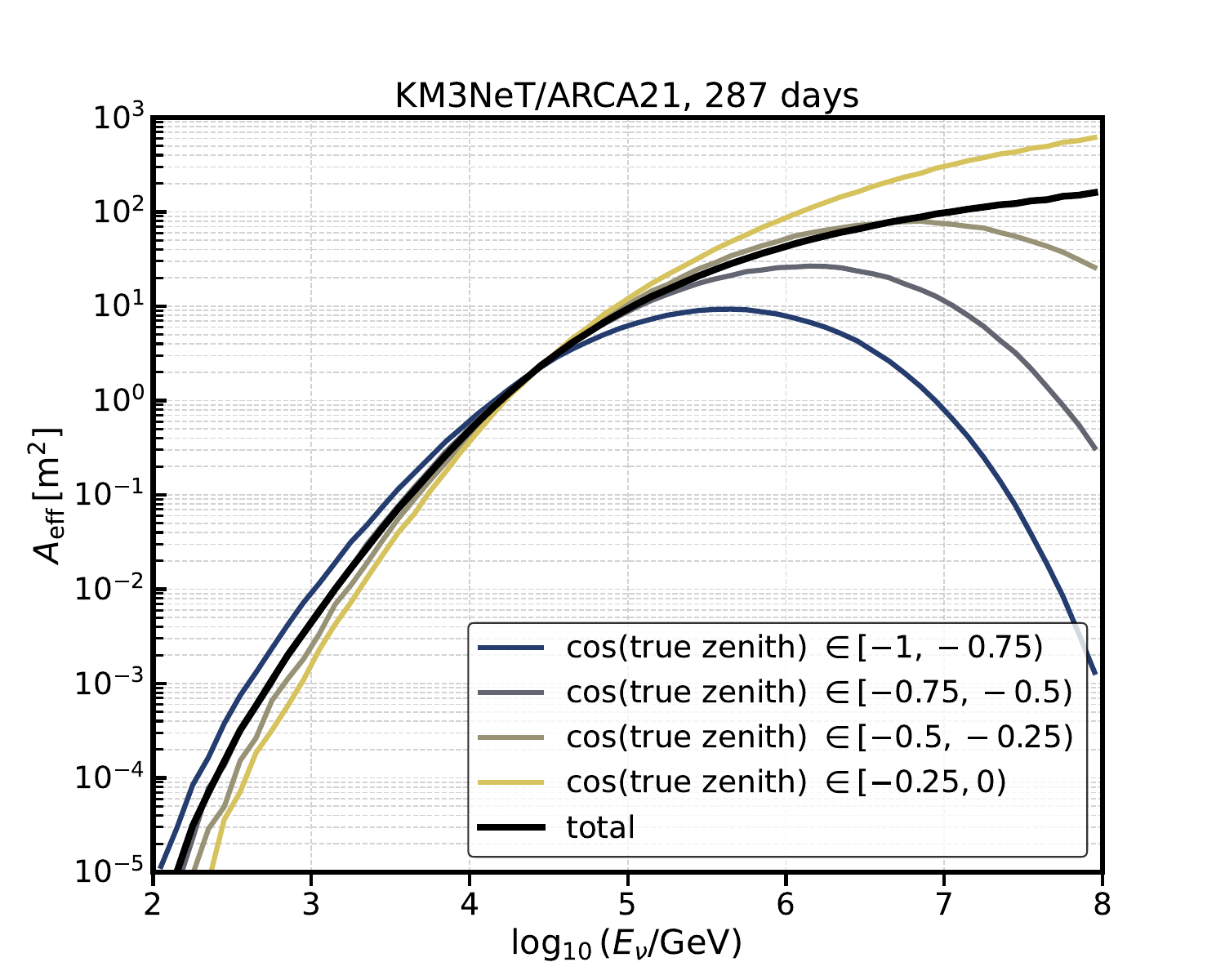}
  \hfill
  \includegraphics[width=0.5\textwidth]{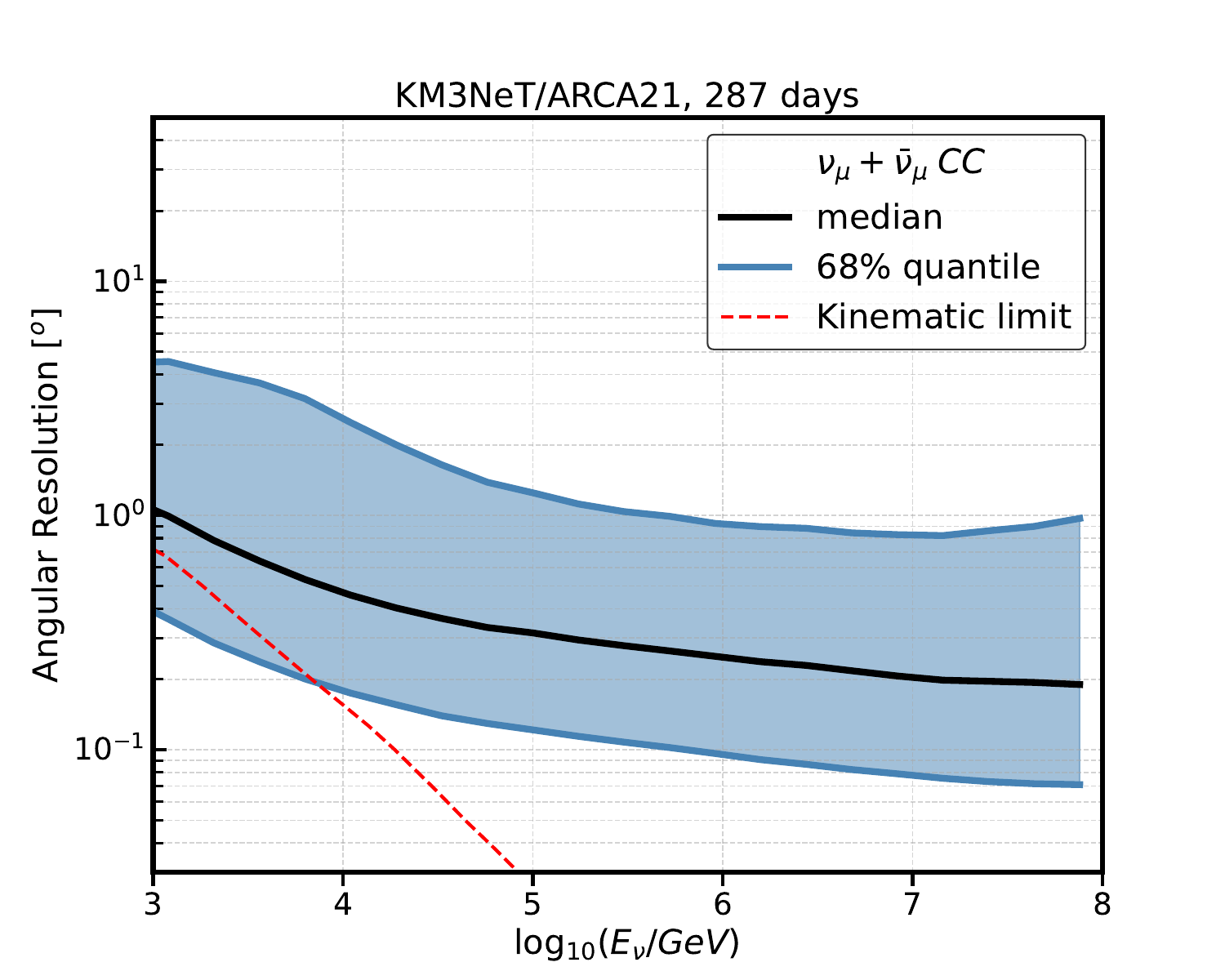}
  \caption{\footnotesize{KM3NeT/ARCA21 effective area, averaged over muon neutrino and anti-neutrino charged current up-going events, for different ranges of true cosine zenith (left). Median (solid black line) and 68\% quantile (shaded blue area) angular resolution for the cosmic muon neutrino charged current channel for ARCA21 as a function of the neutrino energy (right).}}
  \label{fig:mine_effective area}
\end{figure}

The effective area for different zenith angle ranges and the angular resolution for charged-current muon neutrinos as a function of the true neutrino energy are shown in Figure~\ref{fig:mine_effective area} for the ARCA21 detector configuration, considering events selected for the diffuse flux analysis.

%%%%%%%%%%%%%%%%%%%%%%%%%%%%%%%%%%%%%%%%%%%%%%%%%%%%%%%%%%%%%%%%

\subsection{Bayesian analysis results for the all-sky diffuse neutrino flux}\label{sec:All-sky results}
The number of surviving events for the different ARCA detector configurations is reported in Table~\ref{tab:events_after_ecut} and is compared with the expected signal and background events from the MC simulation.

\begin{table}[H]
\centering
%\footnotesize
\caption{\footnotesize Number of events considered in the all-sky diffuse sample for the different ARCA configurations after the BDT and energy cuts reported in Table~\ref{tab:bdt_score_cuts_allsky}. The expected number of signal and background events from the Monte Carlo simulation are given before the best-fit estimation. The number of cosmic neutrinos has been obtained using the flux $\phi_{0} =$ 1.44 $\times$ 10$^{-18}$ GeV$^{-1}$ cm$^{-2}$ s$^{-1}$ sr$^{-1}$ and $\gamma = 2.37$~\cite{IceCube2021_muontracks}.}
\setlength{\tabcolsep}{10pt}
\renewcommand{\arraystretch}{1.4}
\begin{tabular}{l|cccc}%{l|c|c|c|c}
\toprule
\toprule
 & \textbf{ARCA6} & \textbf{ARCA8} & \textbf{ARCA19} & \textbf{ARCA21} \\
\midrule
Atm.~$\nu$ (conv. + prompt) & 11.0 & 17.0 & 7.4 & 24.5 \\
Cosmic~$\nu$ & 0.9 & 2.0 & 0.9 & 4.5 \\
Atm.~$\mu$ & 18.7 & 13.3 & 1.2 & 4.9 \\
Data & 14 & 34 & 11 & 48 \\
\bottomrule
\bottomrule
\end{tabular}
\label{tab:events_after_ecut}
\end{table}

\begin{figure}[H]
  \centering
  \includegraphics[width=0.49\textwidth]{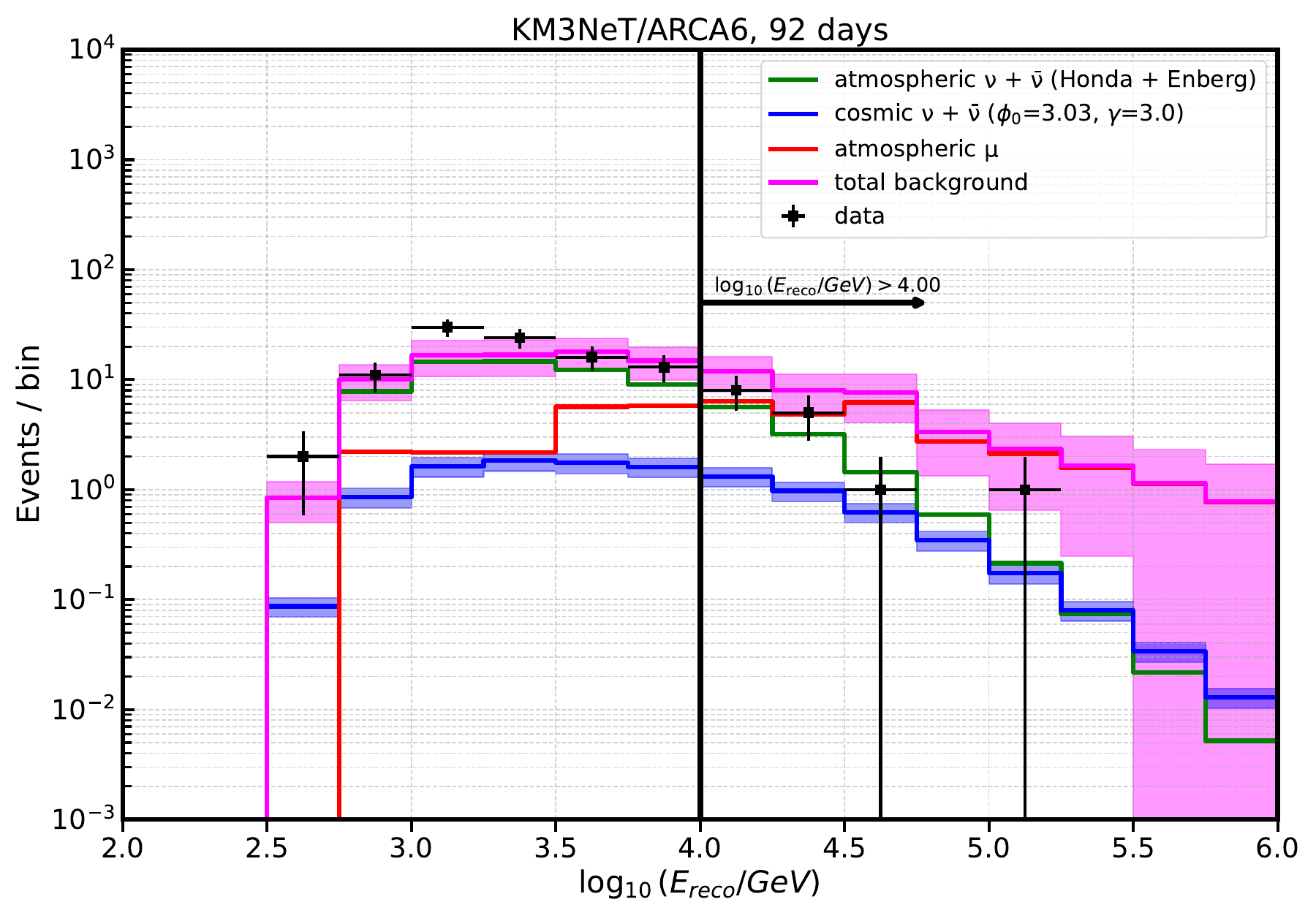}
  \hfill
  \includegraphics[width=0.49\textwidth]{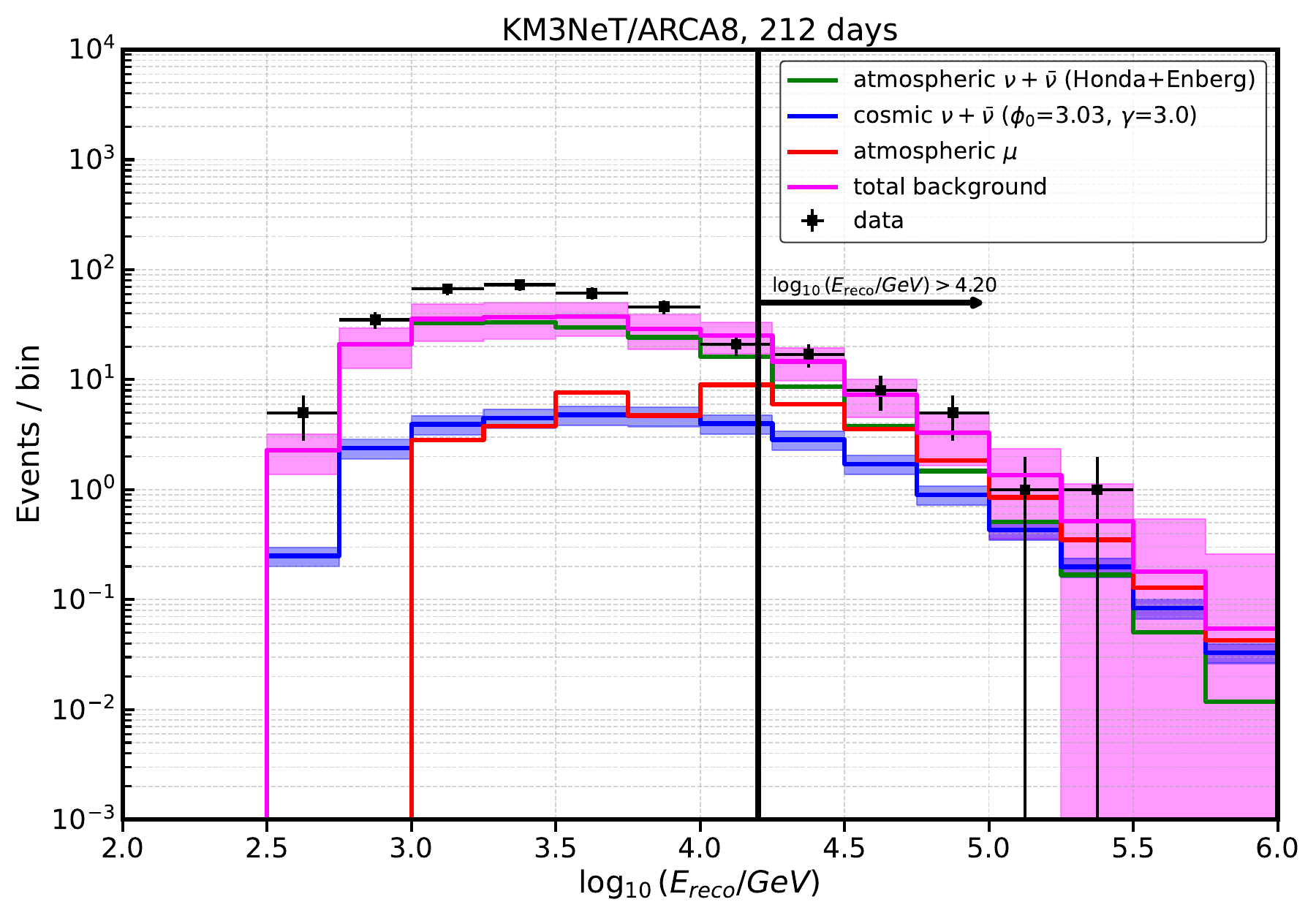}

  \includegraphics[width=0.49\textwidth]{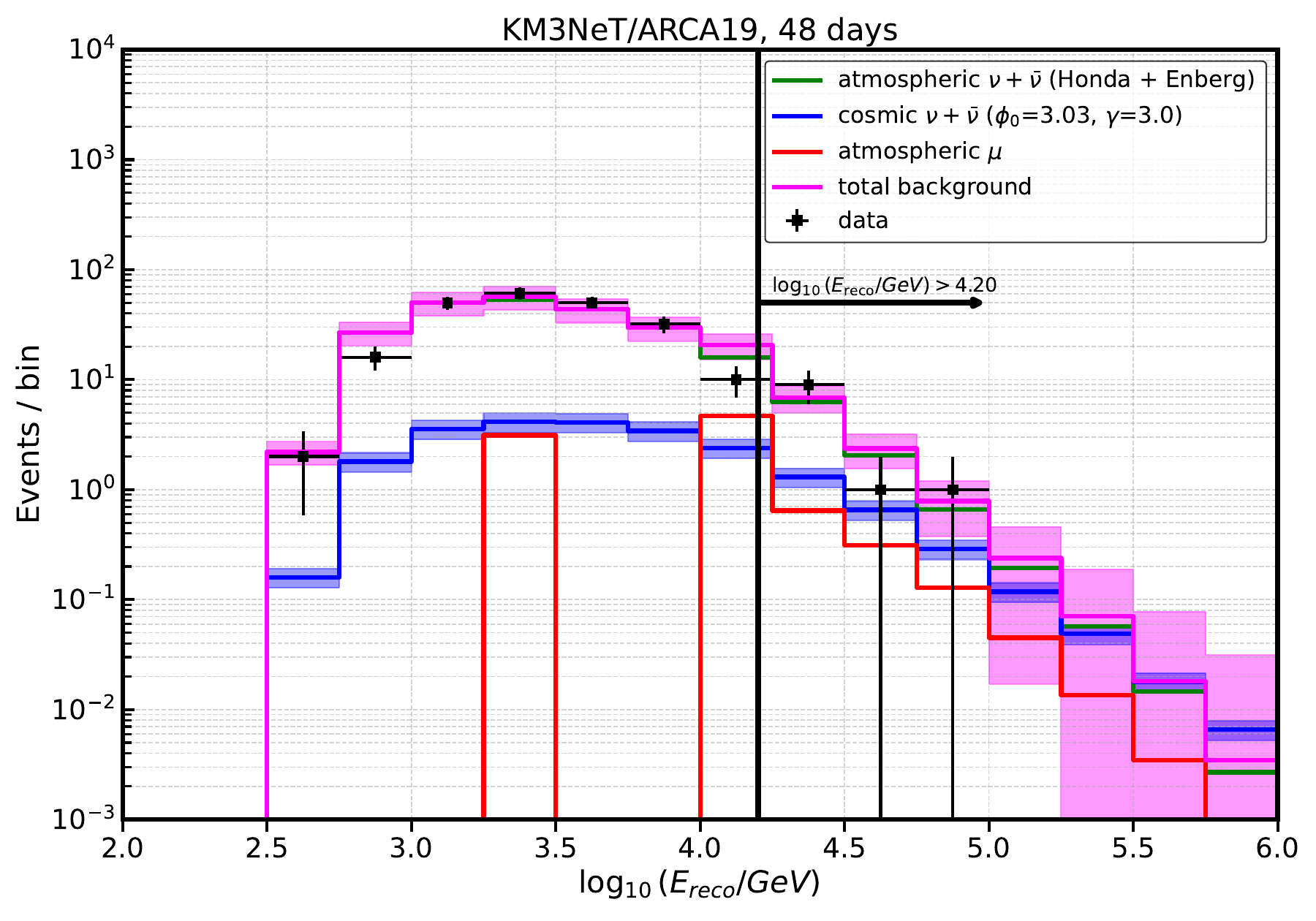}
  \hfill
  \includegraphics[width=0.49\textwidth]{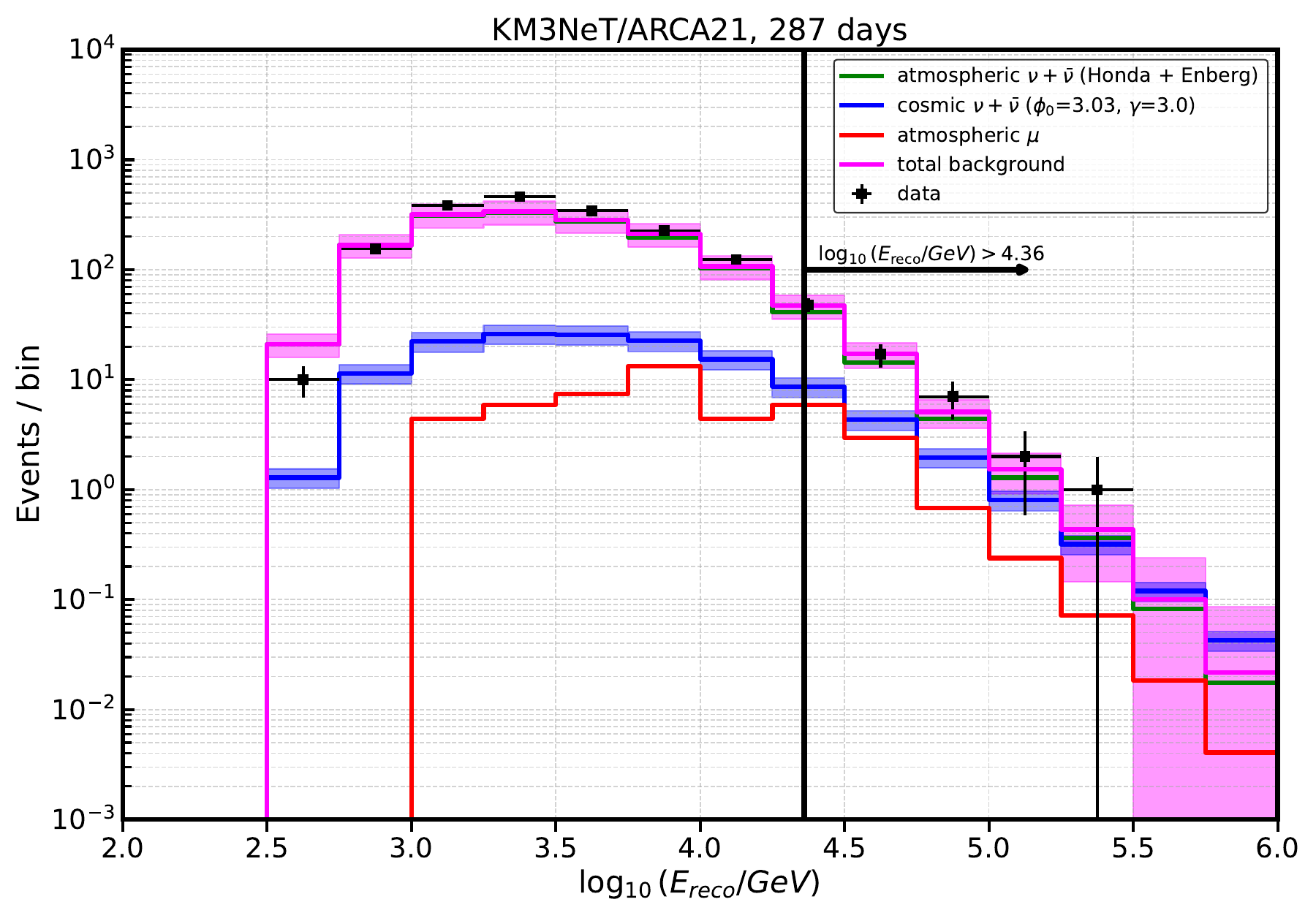}
  \caption{\footnotesize{Distributions of the reconstructed energy using the best-fit estimate for the all-sky cosmic flux parameters for ARCA6$-$21. For ARCA19$-$21 the total background is upscaled by a factor obtained from background normalisation estimation. The expected background of atmospheric neutrinos and muons (purple) and the signal (blue) are shown with their total uncertainty bands as defined in the text. The vertical line in each plot shows the optimal energy threshold cut. The bottom-right panel does not include KM3-230213A~\cite{Nature} because the analysis is selecting strictly up-going events.}}
  \label{fig:afterBDT and after fit LogEreco}
\end{figure}

The resulting energy distributions for all four ARCA configurations are shown in Figure~\ref{fig:afterBDT and after fit LogEreco} assuming the cosmic flux parameters resulting from the fit procedure, with both statistical and systematic uncertainties (see Section~\ref{sec:systematics}) included. 

Among the configurations considered, ARCA21 corresponds to the largest detector geometry analysed and also benefits from the largest livetime, resulting in the dominant event statistics. A comparison between the data collected with the larger geometries (ARCA19 and ARCA21) and the corresponding Monte Carlo simulations suggests that the atmospheric neutrino contribution is underestimated by the simulation. Given the increased statistical power of these configurations, the data from ARCA19 and ARCA21 provide enhanced sensitivity to the overall background modelling and are thus used to constrain the atmospheric component.

For ARCA19 and ARCA21, the background is assumed to fluctuate across energy bins through a single global background normalisation ($N_\text{bkg}$), avoiding the introduction of additional bin-to-bin nuisance parameters that would require substantially larger data samples. Assuming a single background normalisation parameter, its value can be estimated by considering the cosmic parameters $\phi_0$ and $\gamma$ as nuisance parameters, as discussed in Section~\ref{sec:systematics}.
The procedure adopted here optimises therefore the cosmic and atmospheric parameters independently. With higher statistics, a simultaneous fit of all three parameters ($\phi_0,\gamma,N_\text{bkg}$) would be feasible.

\begin{figure}[H]
  \centering
  \includegraphics[width=0.85\textwidth]{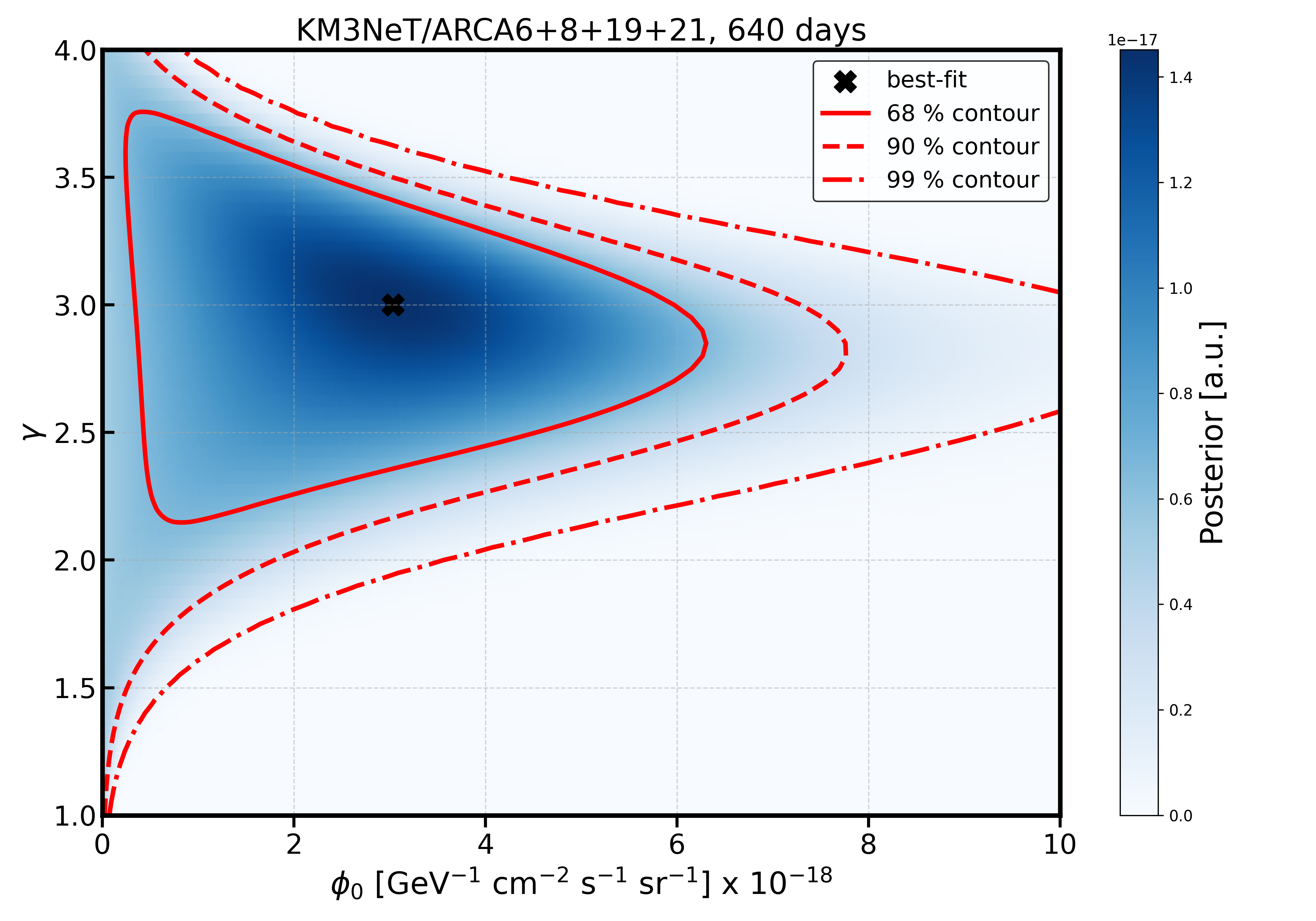}
  \caption{\footnotesize{Marginalised posterior distribution for unblinded data at the final selection stage for the all-sky diffuse flux analysis after the cut on reconstructed energy. The $68\%$, $90\%$ and $99\%$ credible regions for the cosmic parameters together with the best point estimators are shown.}}
  \label{fig:posterior arca6-8-19-21}
\end{figure}

Combining all four periods, the best fit for the cosmic neutrino diffuse flux parameters yields a normalisation of $\phi_0 = 3.0^{+2.1}_{-2.0}$ $\times$ 10$^{-18}$ GeV$^{-1}$ cm$^{-2}$ s$^{-1}$ sr$^{-1}$ following the notation of Eq.~\ref{eq:diffuse flux standard equation} and a spectral index of $\gamma = 3.00^{+0.30}_{-0.35}$, with uncertainties quoted at the $68\%$ Credible Level (C.L.). Credible regions for the cosmic parameters at $68\%$, $90\%$ and $99\%$ together with the best point estimators are shown in Figure~\ref{fig:posterior arca6-8-19-21}.
Given the best-fit parameters above, the expected number of cosmic neutrino events in ARCA6, ARCA8, ARCA19 and ARCA21 is 3.5, 6.8, 2.8 and 11.4, respectively.
The best estimate for the background normalisation is $N_\text{bkg} = 1.53 \pm 0.24$, and in Figure~\ref{fig:afterBDT and after fit LogEreco} the total expected background (atmospheric $\nu$ and $\mu$) in ARCA19 and ARCA21 is upscaled accordingly.

A Pearson correlation analysis was conducted. The results, as shown in~\ref{appendix:correlation}, suggest a moderate negative correlation between $\phi_0$ and the background normalisation, while no significant correlation was observed between spectral index and the other parameters.

\begin{figure}[H]
  \centering
  \includegraphics[width=0.9\textwidth]{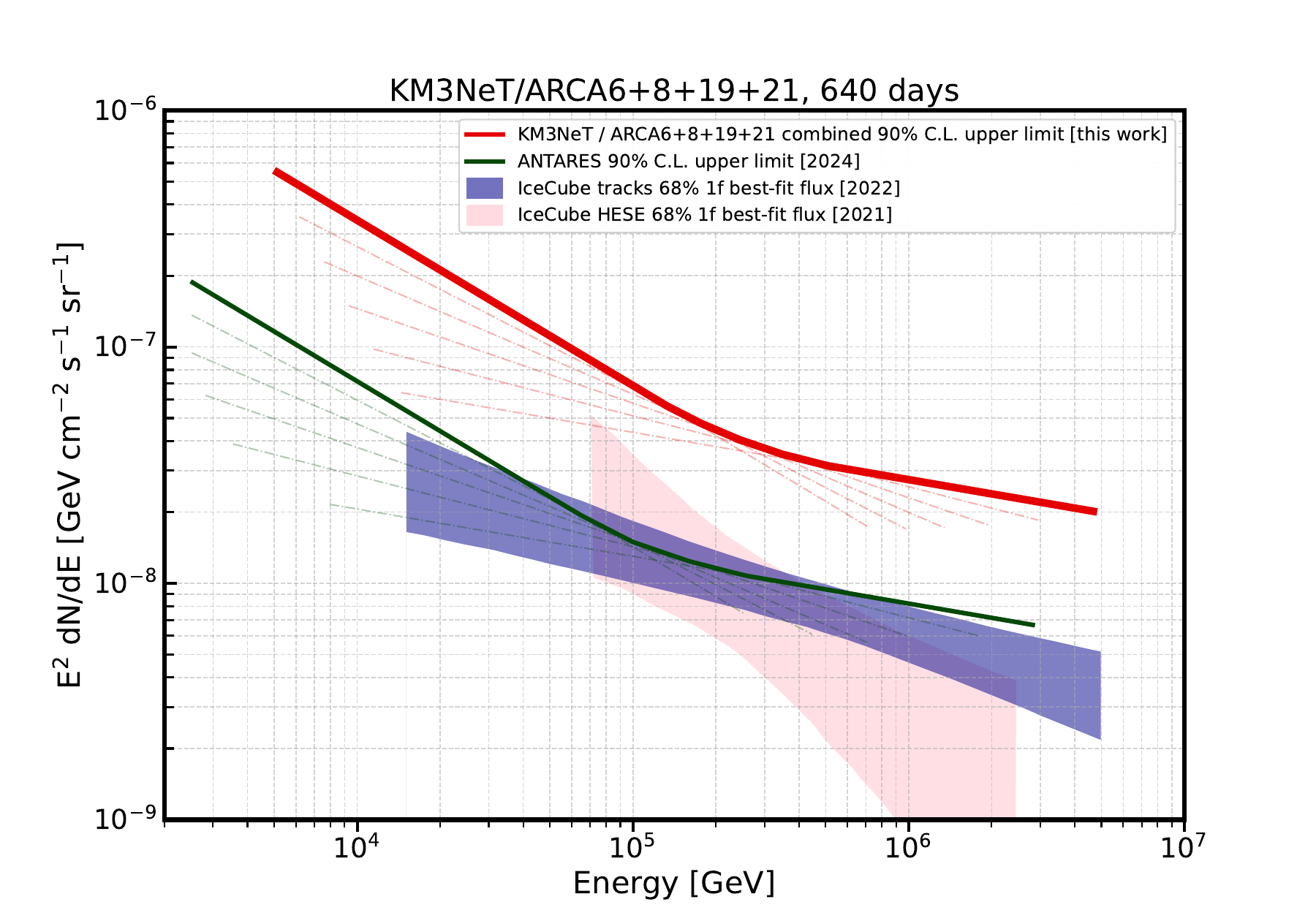}
  \caption{\footnotesize{Envelope of the $90\%$ C.L. upper limits drawn for selected spectral indices $\gamma \in [2.2, 2.7]$ obtained with the full combined ARCA dataset, shown as a function of the true neutrino energy (red line). For each $\gamma$, the line spans the central $90\%$ energy range of the signal. The IceCube~\cite{IceCube2020hese,IceCube2021_muontracks} 68\% C.L. single-flavour best-fit fluxes (pink and blue shaded areas) and the ANTARES~\cite{ANTARES_diffuse_2024} 90\% C.L. upper limits (green line) are included for comparison. }}
  \label{fig:all_arca_sensitivity_ul_btterfly}
\end{figure}

Since the best-fit parameters are not statistically significant, the resulting envelope of the $90\%$ C.L. upper limits for the combined ARCA6 to ARCA21 dataset is shown in Figure~\ref{fig:all_arca_sensitivity_ul_btterfly}. For a direct comparison, the IceCube~\cite{IceCube2020hese,IceCube2021_muontracks} 68\% C.L. single-flavour best-fit fluxes are reported, together with the 90\% C.L. upper limits reported by the ANTARES Collaboration~\cite{ANTARES_diffuse_2024}.

Table~\ref{tab:sensitivities_and_upper_limits_summary} summarises the $90\%$ C.L. sensitivities and upper limits for various spectral indices for the combined ARCA6 to ARCA21 dataset, also indicating the true neutrino energy range containing $90\%$ of the expected signal events.

\begin{table}[H]
\newcolumntype{Y}{>{\centering\arraybackslash}X}
\renewcommand{\arraystretch}{1.5}
\centering
\caption{\footnotesize Sensitivities and upper limits (expressed in units 10$^{-18}$ GeV$^{-1}$ cm$^{-2}$ s$^{-1}$ sr$^{-1}$) at a reference energy $E_{0}$ = 100 TeV at $90\%$ C.L. for different spectral indices $\gamma$ for the combined ARCA6 to ARCA21 dataset. The central $90\%$ energy ranges, expressed in $\log_{10}(E_\nu / \mathrm{GeV})$, define the range for which each measurement is valid.}
\begin{tabularx}{\linewidth}{c|Y|Y|Y}
\toprule
\toprule
\textbf{Spectral Index} & \textbf{Sensitivity} & \textbf{Upper Limit} & \textbf{Energy Range}\\
\hline
%2.0 & 3.03 & 2.81 & [4.37, 7.08] \\
%\hline
%2.1 & 3.82 & 3.55 & [4.26, 6.87] \\
%\hline
2.2 & 4.43 & 4.35 & [4.16, 6.67] \\
%\hline
%2.35 & 5.16 & 5.47 & [4.01, 6.38] \\
2.3 & 4.94 & 5.11 & [4.06, 6.47] \\
%\hline
2.4 & 5.47 &  5.78 & [3.97, 6.29] \\
%\hline
2.5 &   5.89 & 6.31 & [3.88, 6.13] \\
%\hline
2.6 & 6.00 & 6.67 & [3.79, 5.99] \\
%\hline
2.7 & 6.08 & 6.85 & [3.71, 5.85] \\
\bottomrule
\bottomrule
\end{tabularx}
\label{tab:sensitivities_and_upper_limits_summary}
\end{table}

\subsubsection{Frequentist analysis results for the all-sky diffuse neutrino flux}\label{sec:frq_results}

The study for the search of the all-sky diffuse neutrino flux was extended using a frequentist framework. A Poissonian $\chi^2$ is used and systematic uncertainties are incorporated via nuisance parameters with associated pull terms~\cite{fogli2002solar}:

\begin{align}
\label{eq:freq_chi2}
 \chi^2_\alpha  & =  \sum_d\sum_{i} 2 \left( P_{i}^{(d)} - D_{i}^{(d)} +D_{i}^{(d)} ln \frac{D_{i}^{(d)} }{P_{i }^{(d)}} \right) \nonumber \\ 
           &  + \left(  \frac{\alpha_{S}}{\sigma_{S}} \right)^2    
           + \sum_i \left(  \frac{\alpha_{A,i}}{\sigma_{A}} \right)^2 
           + \left(   \frac{\alpha_\text{norm}}{\sigma_\text{norm}}\right)^2 
           + \sum_i  \left(  \frac{\alpha_{M,i}}{\sigma_{M}} \right)^2 ,
\end{align}
where $D_{i}^{(d)}$ refers to the number of selected events in the $i$-th reconstructed energy bin, and the $d$-th detector configuration (ARCA6 to ARCA21). The model prediction $P_{i}^{(d)}$ accounts for the sum of expected atmospheric muons, $M_{i}^{(d)}$, atmospheric neutrinos, $A_{i}^{(d)}$, and cosmic neutrinos, $S_{i}^{(d)}$, as shown in Eq.~\ref{eq:scale}:

\begin{align}
 P_{i}^{(d)} & = (1+\alpha_S) S_{i}^{(d)} + (1+ \alpha_{A,i}  + \alpha_\text{norm} ) A_{i}^{(d)} + (1+ \alpha_{M,i} ) M_{i}^{(d)} ,
 \label{eq:scale}
\end{align}
where $\alpha_S$, $\alpha_{A,i}$, $\alpha_{M,i}$ and $\alpha_\text{norm}$ correspond, respectively, to the systematic uncertainties associated to the signal efficiency, the atmospheric neutrino and atmospheric muon background and the normalisation of the atmospheric neutrino background to account for uncertainties in the atmospheric neutrino flux modelling.

A grid scan over $\phi_0$ and $\gamma$ is performed. For each ($\phi_0$, $\gamma$) pair, the expected signal is computed and $\chi^2$ is minimised with respect to the nuisance parameters.

The best fit found, namely $\phi_0=1.16$ $\times$ 10$^{-18}$ GeV$^{-1}$ cm$^{-2}$ s$^{-1}$ sr$^{-1}$ and $\gamma=3.04$, is mainly driven by ARCA21 due to its larger livetime. The background-only and best-fit curves nearly coincide, indicating a background dominated scenario. The fit quality is $\chi^2$/NDoF = 59.1/46.

\begin{figure}[!ht]
\centering
\includegraphics[width=0.70\textwidth]{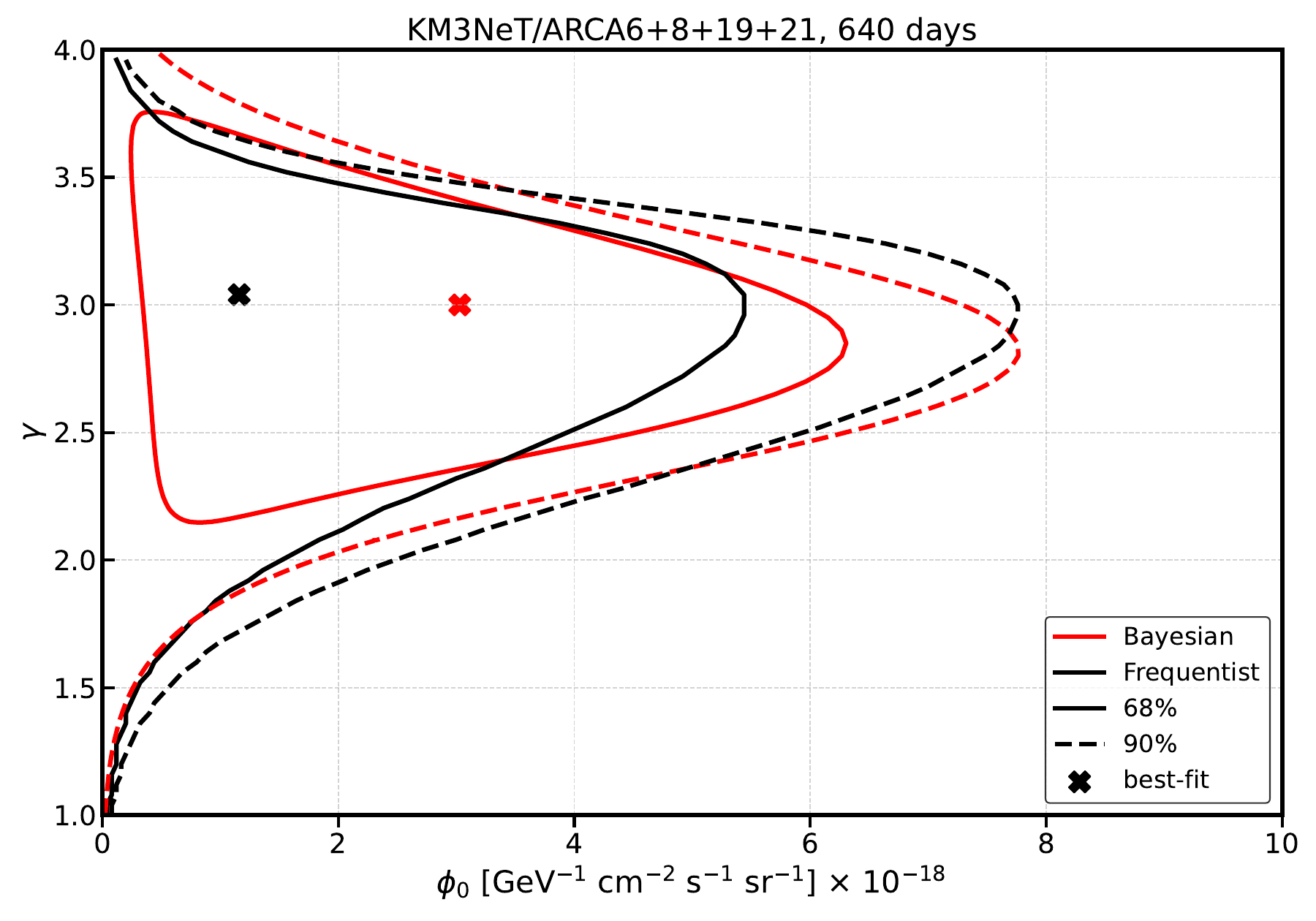} 
\caption{\footnotesize{Comparison between the 68\% (solid) and 90\%  (dashed) confidence-level/credible region contours obtained in the frequentist (black)/Bayesian (red) approach along with the best-fit values (cross) for the all-sky diffuse analysis.}}
\label{fig:freq_ex}
\end{figure}

 The 68\% and 90\% exclusion regions derived from $\Delta\chi^2 = \chi^2 - \chi^2_{\rm min}$ with sharp cuts following~\cite{pdg2022} are shown in Figure~\ref{fig:freq_ex}, overlaid with the corresponding Bayesian 68\% and 90\% contours already reported in Figure~\ref{fig:posterior arca6-8-19-21}.
The Bayesian and frequentist approaches yield results that are consistent within their respective uncertainties.

%%%%%%%%%%%%%%%%%%%%%%%%%%%%%%%%%%%%%%%%%%%%%%%%%%%%%%%%%%%%%%%%%%%%%%%%%%%%%%%%%%%%%%%%%%%
\newpage
\section{Galactic Ridge}\label{sec:GR}

The precise reconstruction of track-like events with an estimated angular resolution better than 0.4$^{\circ}$  for neutrino energies greater than 10 TeV (see Figure \ref{fig:mine_effective area}) provides an exceptional opportunity to observe the region of the Galactic Ridge ($|b| < 2^{\circ}$ and $|l| < 30^{\circ}$ in Galactic coordinates) with unprecedented accuracy and to constrain the spatial and spectral emission of the Galactic diffuse component.
To date, complementary analysis strategies have been employed to search for neutrino emission from the Galactic plane. IceCube \cite{IC_GP} and ANTARES \cite{Theophile} used a template-based likelihood analysis which enables stringent constraints on theoretical models by incorporating predefined spatial and spectral assumptions. This approach, however, is intrinsically dependent on the adopted signal model. To overcome this limitation, a model-independent approach is followed in this work.

\subsection{Analysis method}
The strategy used in this analysis is a model-independent search based on a cut-and-count approach, originally developed by the ANTARES Collaboration \cite{ANTARES_gp}. Event counts and energy distributions are compared in a predefined signal region (\textit{ON region}) to data-driven background estimates obtained from signal-depleted control regions (\textit{OFF regions}).
\begin{figure}[!ht]
	\centering	
	\includegraphics[width=0.8\textwidth]{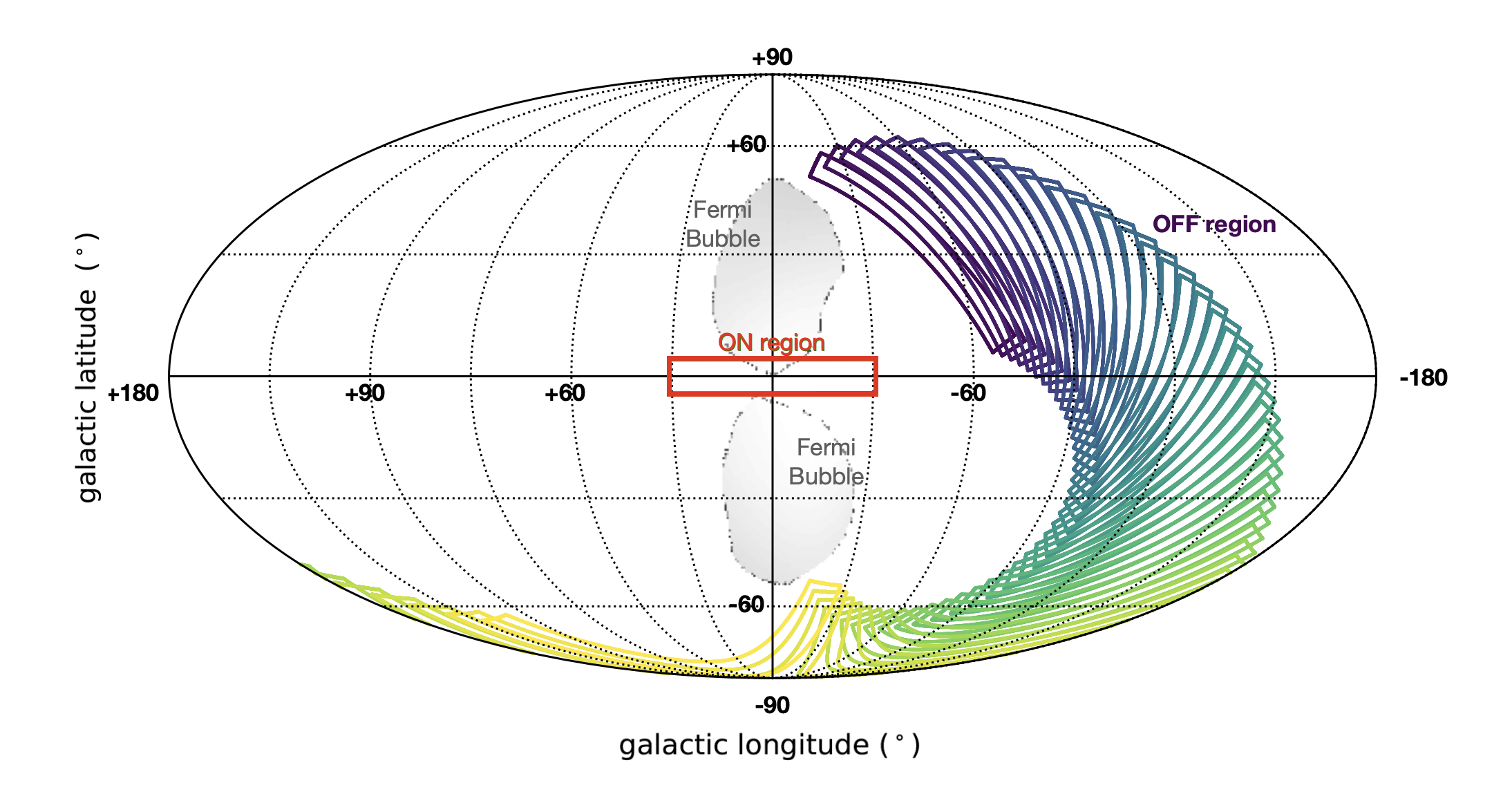}
	\caption{\footnotesize Map in Galactic coordinates of the ON (red rectangle) and OFF (blue to yellow rectangles) regions, corresponding to time-shifted replicas of the same shape.}
	\label{skymap}
\end{figure}
Although the Galactic Ridge is defined by $|b| < 2^{\circ}$ and $|l| < 30^{\circ}$, reconstruction accuracy can cause signal events originating within this region to be reconstructed outside of it. To account for this effect, the ON region is defined using reconstructed coordinates, $|b| < |b_{\mathrm{reco}}|$ and $|l| < |l_{\mathrm{reco}}|$, where these values are the result of the optimisation procedure described in Section~\ref{sec:gr_optimisation}. 
The OFF region is constructed to match the ON region in sky coverage while being shifted in right ascension. For mid-latitude detectors such as KM3NeT, the Earth’s rotation enables observation of different sky regions under equivalent instrumental conditions at different times. In Figure~\ref{skymap}, the ON region is shown as a red rectangle, while OFF regions correspond to time-shifted replicas of the same shape. A continuous OFF region is obtained by applying random time shifts within the interval $[t_0, t_1]$, where $t_0$ and $t_1$ are optimised based on the ON-region extension and constrained to avoid overlap with the Fermi Bubbles. Each event is resampled and weighted according to its probability to fall in the OFF regions after random time shifts.

\subsection{Galactic Ridge event selection}\label{sec:gr_optimisation}
Theoretical models predict significant neutrino emission for the Galactic Ridge starting from energies of approximately 1$-$10 TeV~\cite{CRINGE}. For this reason, an energy cut above 500 GeV is applied, and the selection strategy is optimised starting from the dataset defined in Section~\ref{sec:dataset}. For each of the variables taken into account in the optimisation (reported in Table~\ref{tab:GR_bdt_score_cuts}) different cut values have been considered, linearly spanning a pre-defined range (more details in~\cite{Francesco_thesis}). Therefore, the MRF value is calculated for each possible combination. The optimal point does not depend on the assumed spectrum normalisation, but only on the spectral index, here set to $\gamma$ = 2.4, following the work done in \cite{FermiLAT_2012,ANTARES_gp}.
\begin{table}[H]
\centering
\caption{\footnotesize Optimal set of requirements on BDT score and extension of the Galactic Ridge (reported in Galactic coordinates) applied for each ARCA configuration.}
\begin{tabular}{lccc}
\toprule
\toprule
\textbf{ARCA configuration} & \textbf{BDT Score} & \textbf{$|b_{\text{reco}} (^{\circ})|$} &  \textbf{$|l_{\text{reco}} (^{\circ})|$}\\
\hline
ARCA6  & $> 0.6$ & $<5$ & $<31$ \\
ARCA8  & $> 0.6$ & $<5$ & $<31$ \\
ARCA19 & $> 0.62$ & $<3$ & $<30$ \\ 
ARCA21 & $> 0.62$ & $<3$ & $<30$ \\
\bottomrule
\bottomrule
\end{tabular}
\label{tab:GR_bdt_score_cuts}
\end{table}
After applying the cut on the BDT score, the final sample contains less than 5\% of badly reconstructed atmospheric muons. %, with a signal efficiency exceeding 90\% for events reconstructed with an angular resolution better than 1$^{\circ}$.
%The detector volume enhances the capability to more accurately reconstruct the arrival direction of track-like events and therefore restricting to a smaller region  (as defined for b$_{\text{reco}}$ and l$_{\text{reco}}$ in Table \ref{tab:GR_bdt_score_cuts}), preserving a good signal efficiency.  

\subsection{Galactic Ridge results}\label{sec:Galactic Ridge results}
\begin{figure}[H]
\begin{minipage}[c]{0.5\textwidth}
    \includegraphics[width=1.\textwidth]{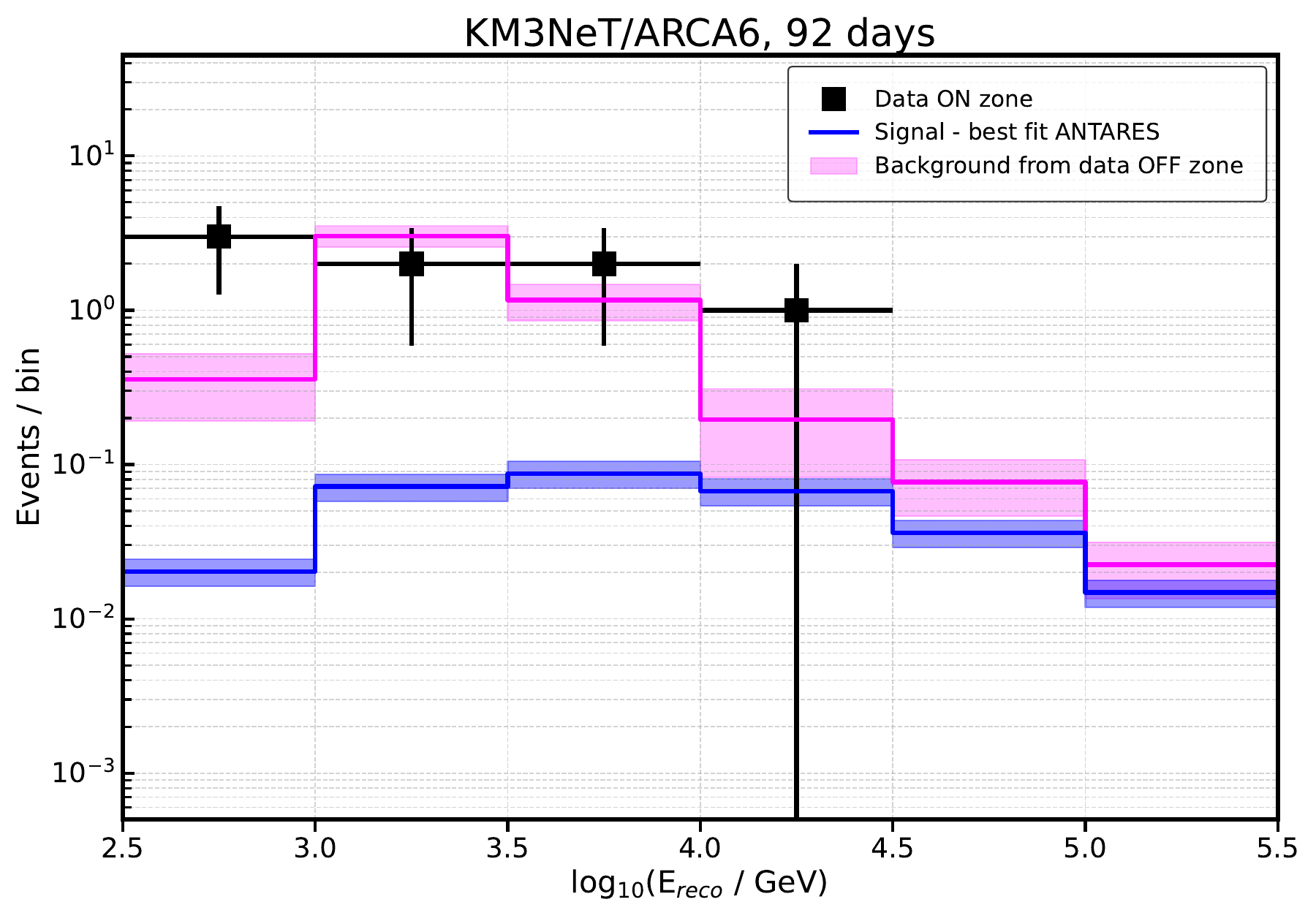}
\end{minipage}
\begin{minipage}[c]{0.5\textwidth}
    \includegraphics[width=1.\textwidth]{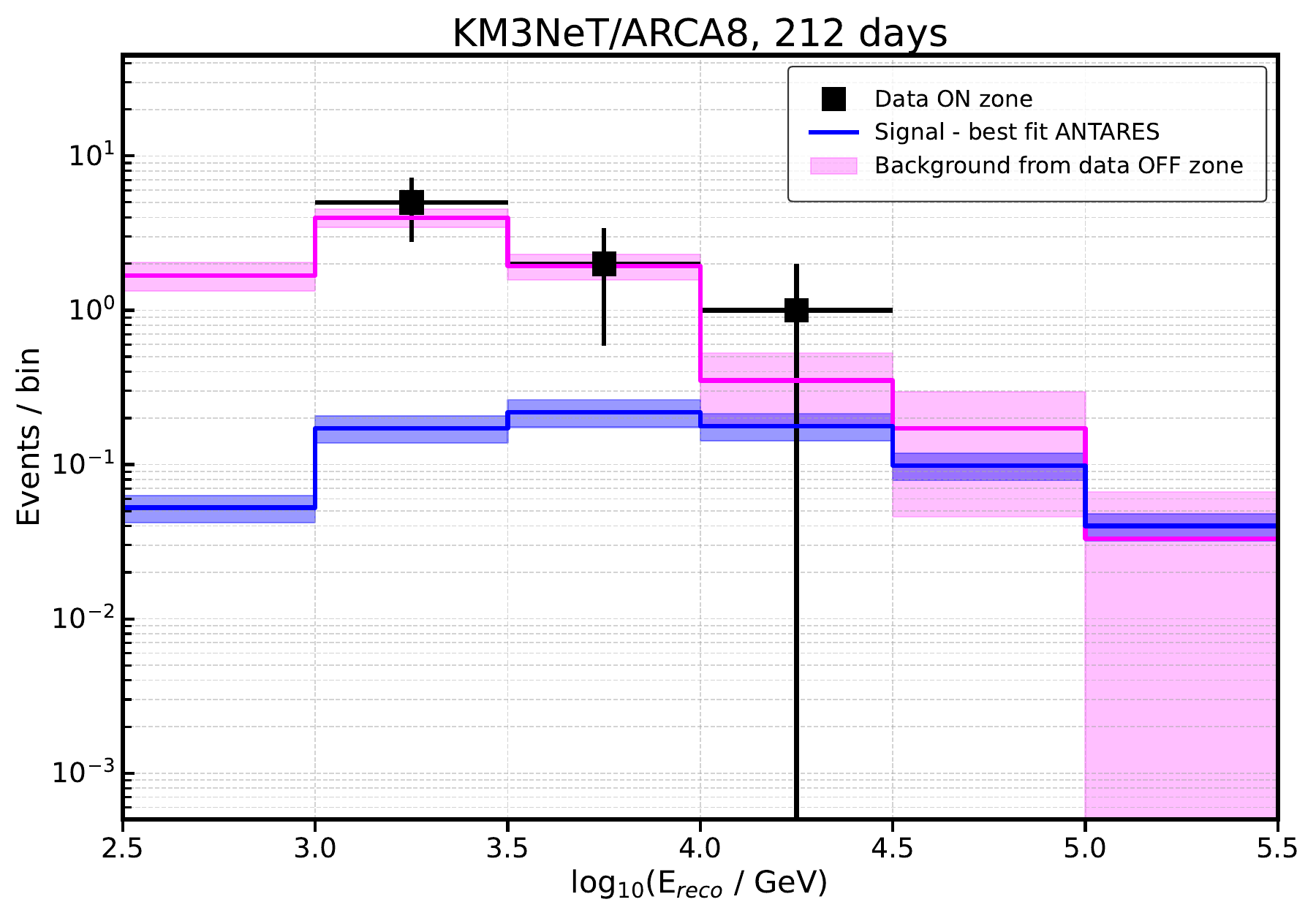}
\end{minipage}
\begin{minipage}[c]{0.5\textwidth}
    \includegraphics[width=1.\textwidth]{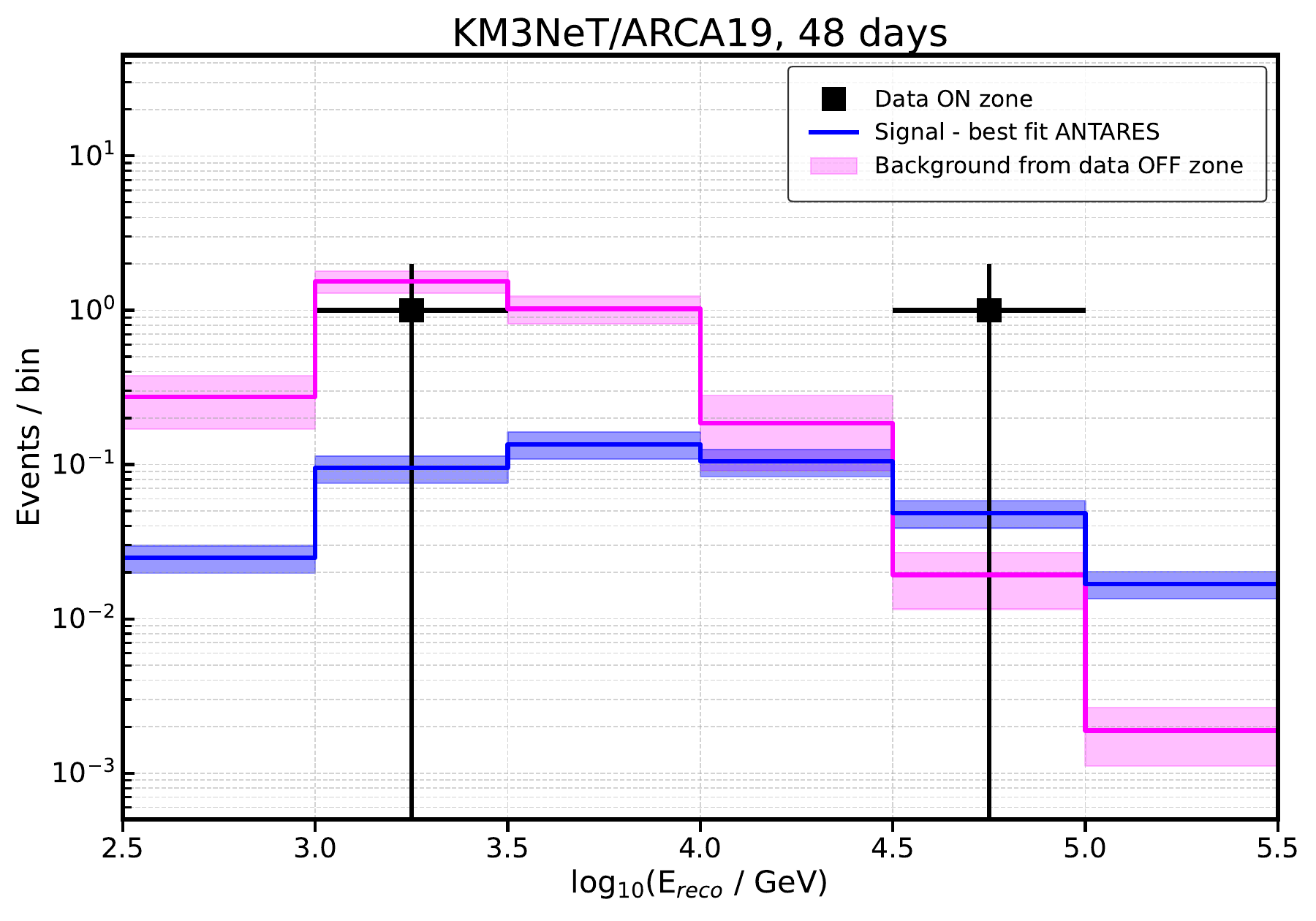}
\end{minipage}
\begin{minipage}[c]{0.5\textwidth}
    \includegraphics[width=1.\textwidth]{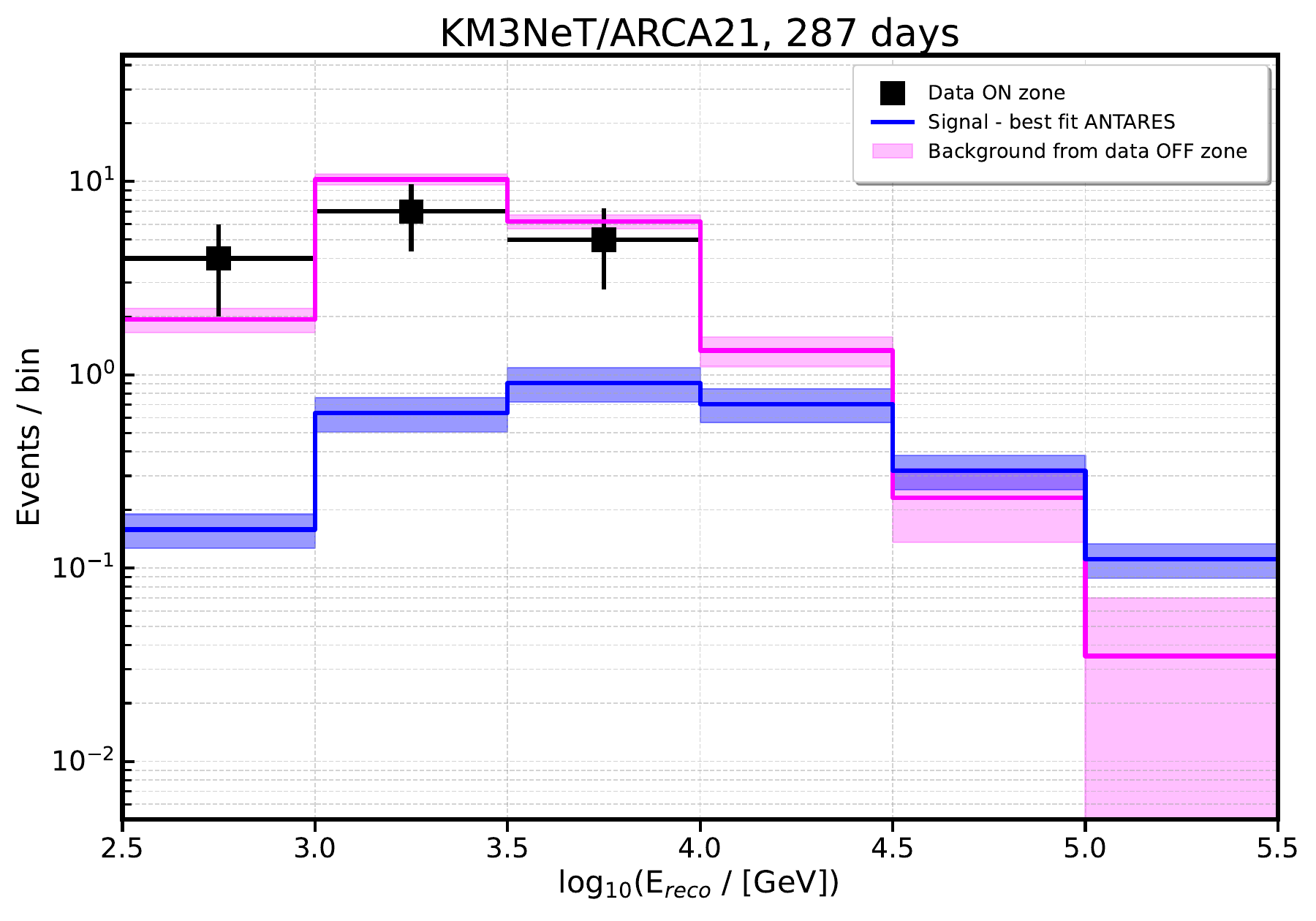}
\end{minipage}
\caption{\footnotesize{Energy distributions for the ARCA6-21 detector configurations. Data points in the Galactic Ridge ON region (black squares) are compared with background estimates derived from OFF regions with the magenta band representing the relative statistical uncertainty. The expected number of events based on the ANTARES best-fit result \cite{ANTARES_gp} is also displayed as a blue line with the associated systematic uncertainty (blue band).}}
\label{unblinded_energy_distributions}
\end{figure}

The energy distribution of surviving events, obtained after unblinding the full data sample, is shown in Figure~\ref{unblinded_energy_distributions} for the various KM3NeT/ARCA detector configurations. Events in the ON region are compared with the expected number of background events. For comparison, the expected number of signal events, assuming the best-fit flux ($\phi_0 = 4.0^{+2.7}_{-2.0}$ $\times$ 10$^{-16}$ GeV$^{-1}$ cm$^{-2}$ s$^{-1}$ sr$^{-1}$ and $\gamma = 2.45^{+0.22}_{-0.34}$, at 40 TeV reference energy) reported in a similar search by the ANTARES Collaboration \cite{ANTARES_gp}, is also displayed.
The energy distributions in the ON region are consistent with the background expectations. The signal flux reported in a similar ANTARES analysis can neither be confirmed nor excluded. The corresponding number of events for ON-region data, background and signal expectations is reported in Table \ref{tab:numbers_GR}.

\begin{table}[H]
\centering
\caption{\footnotesize Number of events found in the Galactic Ridge ON region, background estimates using data in the OFF region, and signal events expected assuming the best-fit flux of ANTARES are given for the ARCA6-21 detector configurations.}
\begin{tabular}{lcccc}
\toprule
\toprule
\makecell[l]{\textbf{ARCA configuration}} & \makecell{\textbf{ON region}\\\textbf{events}} & \makecell{\textbf{Background from}\\\textbf{data OFF zone}} & \makecell{\textbf{Expected signal}\\\textbf{ANTARES best-fit}}\\
\midrule
 ARCA6 & 8 & 4.8 & 0.3 \\ 
 ARCA8 & 8 & 8.1  & 0.8 \\
 ARCA19 & 2 & 3.0 & 0.4   \\
 ARCA21 & 16 & 19.9 & 2.9  \\
\bottomrule
\bottomrule
\end{tabular}
\label{tab:numbers_GR}
\end{table}

\begin{figure}[H]
	\centering	
	\includegraphics[width=0.9\textwidth]{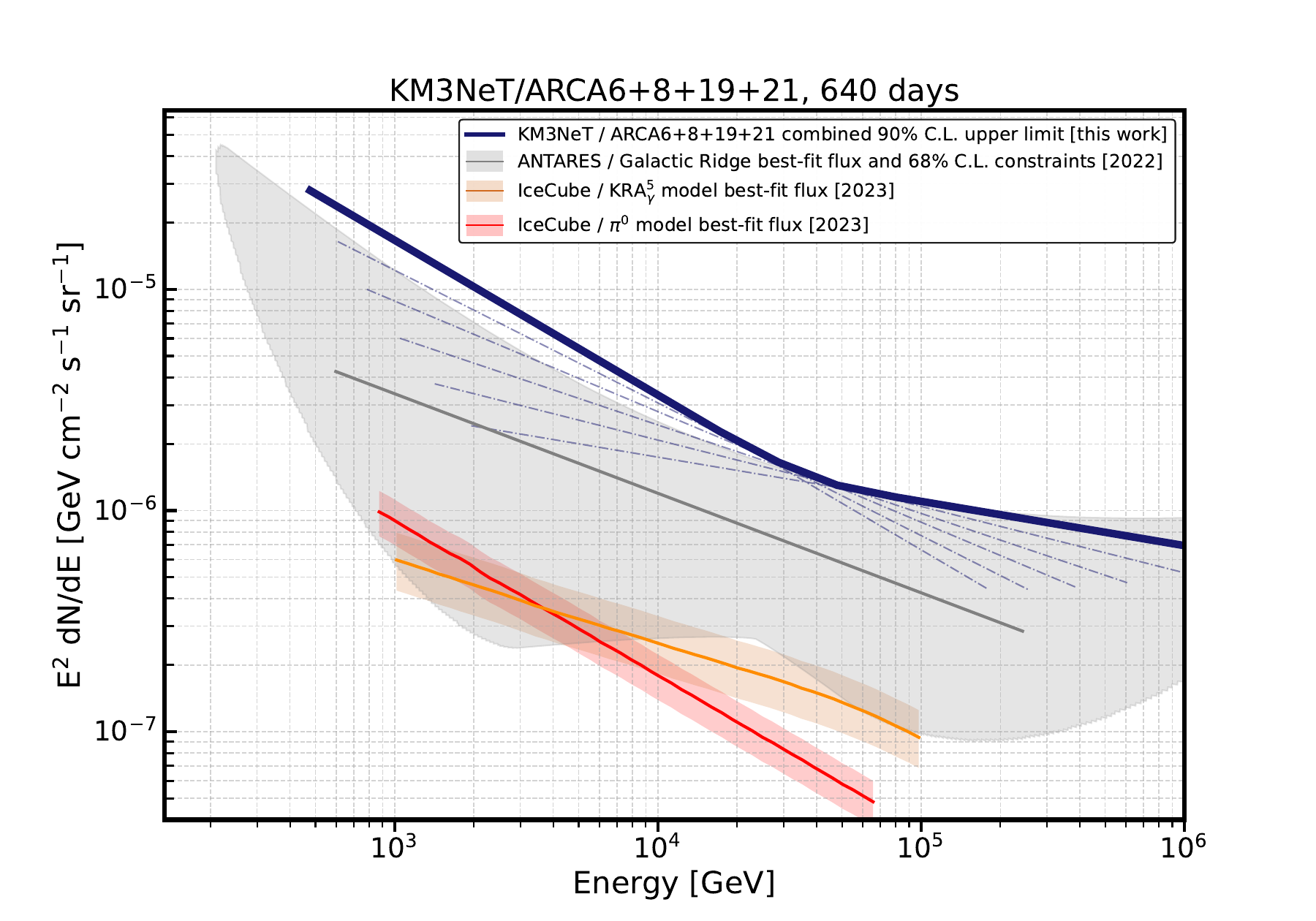}
	\caption{\footnotesize Envelope of 90$\%$ C.L. upper limit for $\gamma \in [2.2, 2.7]$ for the Galactic Ridge diffuse flux shown as a function of the true neutrino energy (blue line). For comparison, ANTARES and IceCube best-fit fluxes are reported. IceCube limits have been rescaled considering the signal fraction contained in the Galactic Ridge. }
	\label{fig:ul_ridge}
\end{figure}

The Bayesian method described in Section~\ref{sec:systematics} has been applied. Given the use of a data-driven background estimate, only the systematic uncertainty on the signal acceptance is taken into account.
Derived sensitivities and ULs, computed for the central 90\% energy range of the signal events for spectral indices in the range [2.2, 2.7]  are reported in Table~\ref{tab:GR_final_UL}. 
\begin{table}[H]
\newcolumntype{Y}{>{\centering\arraybackslash}X}
\renewcommand{\arraystretch}{1.5}
\centering
\caption{\footnotesize 90$\%$ C.L. sensitivity and upper limits for a Galactic Ridge diffuse flux are given for the combined ARCA6 to 21 dataset. The quoted values correspond to a single power-law assumption and for spectral indices $\gamma$ ranging from 2.2 to 2.7. All results are expressed in units of GeV$^{-1}$ cm$^{-2}$ s$^{-1}$ sr$^{-1}$ at a reference energy $E_0$ = 1 GeV. The central $90\%$ energy ranges, expressed in $\log_{10}(E_\nu / \mathrm{GeV})$, define the domain over which each measurement is valid.}
\label{tab:GR_final_UL}
\begin{tabularx}{\linewidth}{c|Y|Y|Y}
\toprule
\toprule
\textbf{Spectral Index} & \textbf{Sensitivity} & \textbf{Upper Limit} & \textbf{Energy Range}\\
\hline
2.2 &  9.5 $\cdot10^{-6}$ &  1.1 $\cdot10^{-5}$ & [3.25, 6.18] \\
%\hline
2.3 & 2.9 $\cdot10^{-5}$ &  3.3 $\cdot10^{-5}$ & [3.11, 5.94] \\
%\hline
2.4 & 8.7 $\cdot10^{-5}$ &  9.7 $\cdot10^{-5}$ & [2.98, 5.73] \\
%\hline
2.5 &   2.6 $\cdot10^{-4}$ & 2.8 $\cdot10^{-4}$ & [2.86, 5.54] \\
%\hline
2.6 & 7.2 $\cdot10^{-4}$ & 7.7 $\cdot10^{-4}$ & [2.75, 5.36] \\
%\hline
2.7 & 1.9 $\cdot10^{-3}$ & 2.1 $\cdot10^{-3}$ &  [2.65, 5.20]\\
\bottomrule
\bottomrule
\end{tabularx}
\end{table}
In the absence of a statistically significant excess of events originating from the Galactic Ridge region, the envelope of upper limits is presented in Figure~\ref{fig:ul_ridge} and is compared to ANTARES~\cite{ANTARES_gp} and IceCube~\cite{IC_GP} results.
The IceCube limits, derived using a template-fitting analysis approach, have been rescaled to account for the fraction of the template signal contained within the Galactic Ridge region. Nevertheless, this comparison should be interpreted with caution in light of the differences in the underlying analysis methodologies.

%%%%%%%%%%%%%%%%%%%%%%%%%%%%%%%%%%%%%%%%%%%%%%%%%%%%%%%%%%%%%%%%%%%%%%%%%%%%%%%%%%%%%%%%%%%

\section{Conclusion and outlook}
\label{sec:conclusions}
This paper presents the first search for a diffuse astrophysical neutrino flux from both the full sky and the Galactic Ridge region using KM3NeT/ARCA data from the configurations ARCA6, ARCA8, ARCA19, and ARCA21, corresponding to a total effective livetime of 640 days.

No statistically significant excess of cosmic neutrino events above the estimated background is observed in either search. Fitting for the all-sky single-flavour cosmic neutrino flux parameters yields a normalisation of $\phi_0^{1f} = 3.0^{+2.1}_{-2.0} \times 10^{-18}$ GeV$^{-1}$ cm$^{-2}$ s$^{-1}$ sr$^{-1}$, a spectral index of $\gamma = 3.00^{+0.30}_{-0.35}$, and an atmospheric background normalisation consistent with 1.53 at the 68\% credible level.
The computed ULs are fully compatible with previous measurements reported by other experiments. At the same time, the analysed dataset does not yet provide sufficient sensitivity to detect a signal from the Galactic Ridge region. Since the 68\% credible level does not provide a well-constrained solution for the Galactic Ridge fit, no flux or spectral parameters are reported for this search. Although no significant signal is observed, these analyses demonstrate the capability of KM3NeT/ARCA to yield competitive diffuse flux measurements even with partial detector configurations. 
The expected increase in exposure, driven by both longer livetime and further detector expansion, will significantly enhance the sensitivity to both all-sky and spatially localised emission, particularly in the southern sky and along the Galactic plane.

%%%%%%%%%%%%%%%%%%%%%%%%%%%%%%%%%%%%%%%%%%%%%%%%%%%%%%%%%%%%%%%%%%%%%%%%%%%%%%%%%%%%%%%%%%%

\section{Acknowledgements}
The authors acknowledge the financial support of:
%INFRADEV
KM3NeT-INFRADEV2 project, funded by the European Union Horizon Europe Research and Innovation Programme under grant agreement No 101079679;
%Belgium
Funds for Scientific Research (FRS-FNRS), Francqui foundation, BAEF foundation.
%Czeck
Czech Science Foundation (GAČR 24-12702S);
%France
Agence Nationale de la Recherche (contract ANR-15-CE31-0020), Centre National de la Recherche Scientifique (CNRS), Commission Europ\'eenne (FEDER fund and Marie Curie Program), LabEx UnivEarthS (ANR-10-LABX-0023 and ANR-18-IDEX-0001), Paris \^Ile-de-France Region, Normandy Region (Alpha, Blue-waves and Neptune), France;
The ACME project funded by the European Union’s Horizon Europe Research and innovation programme under Grant Agreement No 101131928;
%For the CPER
The Provence-Alpes-Côte d'Azur Delegation for Research and Innovation (DRARI), the Provence-Alpes-Côte d'Azur region, the Bouches-du-Rhône Departmental Council, the Metropolis of Aix-Marseille Provence and the City of Marseille through the CPER 2021-2027 NEUMED project,
%For IN2P3
The CNRS Institut National de Physique Nucléaire et de Physique des Particules (IN2P3);
%Georgia
Shota Rustaveli National Science Foundation of Georgia (SRNSFG, FR-22-13708), Georgia;
%Germany (Max Planck Inst.) 
ERC MuSES project No 101142396); 
%Germany Wurzburg
ERC starting grant MessMapp, under contract No. 949555;
%Greece
The General Secretariat of Research and Innovation (GSRI), Greece;
V. Tsourapis acknowledges the support of the Hellenic Foundation
for Research and Innovation (HFRI) under the 3rd Call for HFRI PhD Fellowships (Fellowship Number: 5403).
%Italy
Istituto Nazionale di Fisica Nucleare (INFN) and Ministero dell’Universit{\`a} e della Ricerca (MUR). KM3NeT4RR MUR Project National Recovery and Resilience Plan (NRRP), Mission 4 Component 2 Investment 3.1, Funded by the European Union – NextGenerationEU,CUP I57G21000040001, Concession Decree MUR No. n. Prot. 123 del 21/06/2022;
%Morocco
Ministry of Higher Education, Scientific Research and Innovation, Morocco, and the Arab Fund for Economic and Social Development, Kuwait;
%The Netherlands
Nederlandse organisatie voor Wetenschappelijk Onderzoek (NWO), the Netherlands;
%Poland
The grant “AstroCeNT: Particle Astrophysics Science and Technology Centre”, carried out within the International Research Agendas programme of the Foundation for Polish Science financed by the European Union under the European Regional Development Fund; The program: “Excellence initiative-research university” for the AGH University in Krakow; The ARTIQ project: UMO-2021/01/2/ST6/00004 and ARTIQ/0004/2021;
%Romania
Ministry of Education and Scientific Research, Romania
%Slovak Republic
Slovak Research and Development Agency under Contract No. APVV-22-0413; Ministry of Education, Research, Development and Youth of the Slovak Republic;
%Spain 
MICIU for PID2024-156285NB-C41, -C42- C43, funded by MICIU/AEI/10.13039/501100011033 and by FEDER, EU, and for CNS2023-144099; Generalitat Valenciana for CIDEGENT/2020/049, CIDEGENT/2021/23, CIDEIG/2023/20, CIPROM/2023/51 and INNVA1/2024/110 (IVACE+i), and Fundaci\'{o}n Bancaria La Caixa (ID 100010434), for LCF/BQ/PI25/12100025, Spain;
%UAE
Khalifa University internal grants (ESIG-2023-008, RIG-2023-070 and RIG-2024-047), United Arab Emirates;
%UK
The European Union's Horizon 2020 Research and Innovation Programme (ChETEC-INFRA - Project no. 101008324).
% disclaimer
Views and opinions expressed are those of the author(s) only and do not necessarily reflect those of the European Union or the European Research Council. Neither the European Union nor the granting authority can be held responsible for them.

\newpage
\bibliographystyle{elsarticle-num}
\bibliography{cleaned_bibliography_natbib}
\newpage
\input{Appendix}

\end{document}

%% file: Appendix.tex
\appendix
%\section{Appendix}
%\addcontentsline{toc}{section}{Appendix}

\section{BDT architecture}
\label{appendix:bdt}
The optimised BDT configuration consists of: Number of trees (\texttt{NTrees}): \textbf{1200}; Maximum tree depth (\texttt{MaxDepth}): \textbf{6}; Minimum node size (\texttt{MinNodeSize}): \textbf{1}; Boosting parameter (\texttt{AdaBoostBeta}): \textbf{0.02}; Number of grid points in variable splitting (\texttt{NCuts}): \textbf{14}. For the sake of simplicity, the models applied on ARCA6-8 and ARCA19-21 were ultimately configured to share the same architecture. 

The variables used for the BDT training are reported in Table~\ref{tab:variables_importance}.

\begin{table}[H]
\newcolumntype{Y}{>{\centering\arraybackslash}X}
\renewcommand{\arraystretch}{1.5}
\centering
\caption{\footnotesize{Selected input variables used to train the BDT models.}}
\begin{tabularx}{\linewidth}{Ycc}
\toprule
\toprule
\textbf{Variable} \\
\hline
Angular error of the track reconstruction parameters \\
Length of track between the vertex and last emitted photon using the Cherenkov hypothesis \\
$z$ direction from the shower reconstruction \\
Fraction of triggered hits on lower hemispheres \\
Likelihood of the reconstructed tracks \\
Fraction of early hits in time coincidence with track, when the distance between the vertex and photon emission point is more than 10 m. \\
Number of good solutions. A solution is good if the difference in angle between best solution and current is less than 1$^\circ$ \\
max $\lbrace$angle between best track and a solution$\rbrace$ \\
$z$ vertex position from the track reconstruction \\
Difference between the maximum likelihood of the up-going track reconstruction solutions and the maximum likelihood of the down-going track reconstruction solutions over the likelihood of best track \\
Number of DOMs respecting the Cherenkov condition \\
Minimum zenith amongst the track reconstruction solutions \\
Number of triggered DUs \\
Number of downgoing solutions \\
max$\lbrace$Time over Threshold (ToT) of triggered hits of the event$\rbrace$ \\
\bottomrule
\bottomrule
\end{tabularx}
\label{tab:variables_importance}
\end{table}

\section{Correlation matrices}
\label{appendix:correlation}
\begin{figure}[h]
  \centering
  \includegraphics[width=0.5\textwidth]{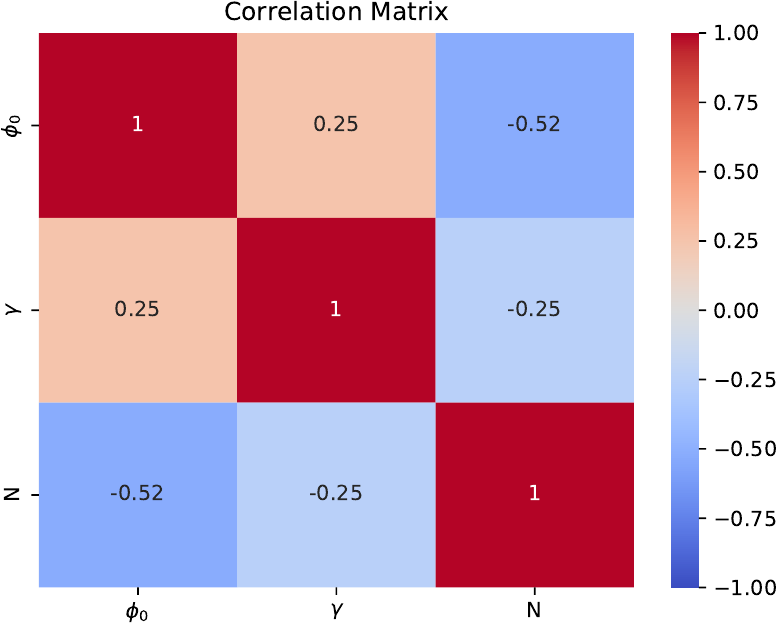}
  \caption{\footnotesize{Pearson correlation matrix for the parameters $\text{N}_{\rm bkg}$, $\phi_0$, and $\gamma$ for the combined ARCA19+21 configurations.}}
  \label{fig:Pearson_arca_19_21}
\end{figure}

\begin{figure}[H]
  \centering
  \includegraphics[width=0.7\textwidth]{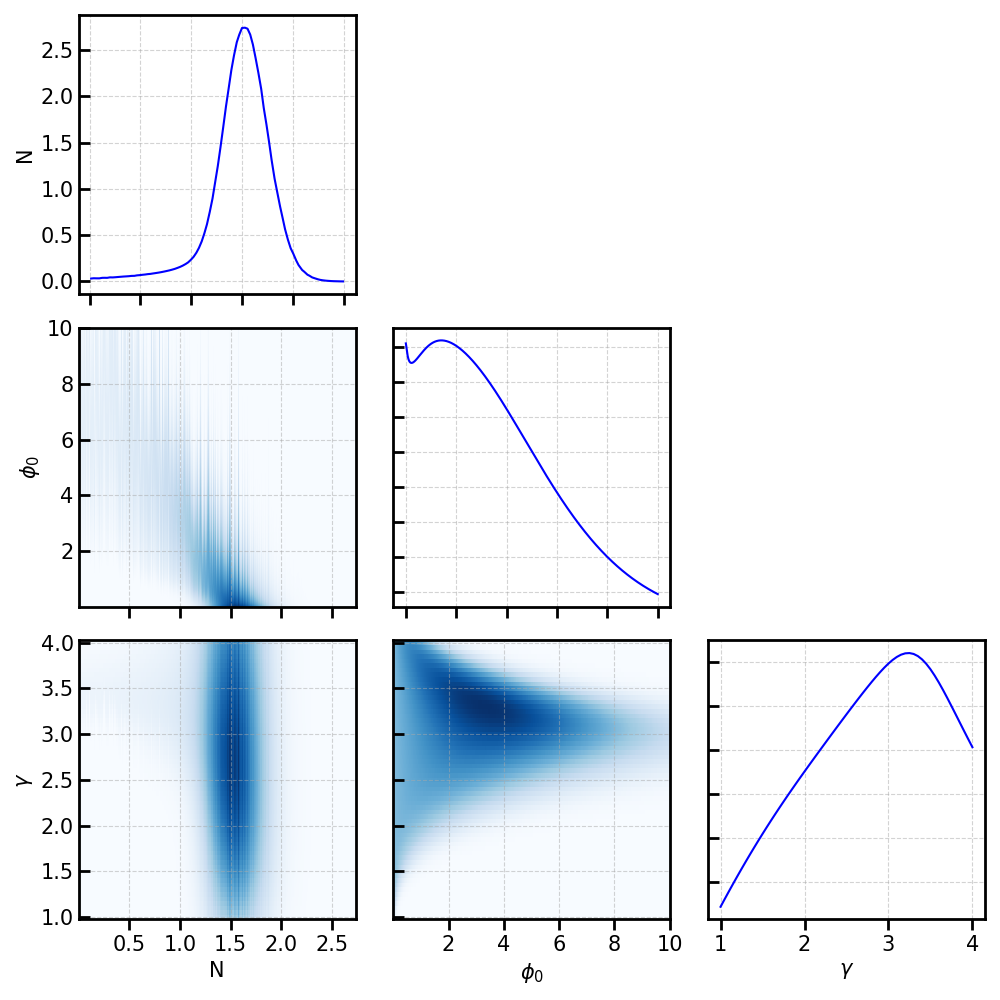}
  \caption{\footnotesize{Corner plot for the parameters $\text{N}_{\rm bkg}$, $\phi_0$, and $\gamma$ for the combined ARCA19+21 configurations.}}
  \label{fig:seabornlike}
\end{figure}

The Pearson correlation matrix summarises linear correlations between variables, with coefficients spanning from -1 (fully anti-correlated) to +1 (fully correlated). It offers a compact way to assess how parameters are interrelated and whether degeneracies may exist in the analysis. The Pearson correlation matrix for the ARCA19+21 combination is shown in Figure~\ref{fig:Pearson_arca_19_21}. A corner plot of the three parameters is given in Figure~\ref{fig:seabornlike}.

